\documentclass[12pt]{article}

\usepackage{epsf,epsfig}
\usepackage{graphics}
\usepackage{bbold}
\usepackage{dsfont}
\usepackage{textcomp}
\usepackage{comment}
\usepackage{cite}
\usepackage{slashed}
\usepackage{amsmath,amssymb,booktabs}
\def\be{\begin{equation}}
\def\ee{\end{equation}}
\def\bea{\begin{eqnarray}}
\def\eea{\end{eqnarray}}

\def\nn{\nonumber}

\begin{document}
\thispagestyle{empty}
\allowdisplaybreaks

\begin{flushright}
{\small
TUM-HEP-943/14\\
OUTP-14-08P\\
May 20, 2014}
\end{flushright}

\vspace{0.8cm}
\begin{center}
\Large\bf\boldmath
 $(g-2)_\mu$ in the custodially protected RS model
\unboldmath
\end{center}

\vspace{1cm}
\begin{center}
{\sc P.~Moch$^{a}$} and 
{\sc J.~Rohrwild$^{b}$}\\[5mm]
{\sl ${}^a$Physik Department T31,\\
James-Franck-Stra\ss e~1, Technische Universit\"at M\"unchen,\\
D--85748 Garching, Germany}\\[5mm]
{\sl ${}^b$Rudolf Peierls Centre for Theoretical Physics,\\ University of Oxford,
1 Keble Road,\\ Oxford OX1 3NP, United Kingdom}\\[1.5cm]
\end{center}

\vspace{1.3cm}
\begin{abstract}
\vspace{0.2cm}\noindent
We study leptonic operators of dimension six in the custodially protected Randall-Sundrum model. 
Their contribution to the anomalous magnetic moment of 
the muon is evaluated. We find that the contribution to g-2
due to diagrams with an internal gauge boson exchange 
is given by 
$$\Delta a_\mu^{\rm g}\approx 2.72 \times 10^{-10}\left(\frac{1 \;\rm TeV}{T}\right)^2  $$
basically independent of the model parameters except for the KK mass scale 
T---a factor of more than $3$ larger than in a scenario without 
custodial protection. We also investigate the impact of contributions to
the dipole operators due to an internal Higgs exchange, which can provide  
sizable albeit model-dependent corrections.

\end{abstract}

\newpage
\setcounter{page}{1}


\section{Introduction}

The anomalous magnetic moment of the muon is one of the 
most precisely predicted and measured observables in particle physics. Over the
last 65 years tremendous effort has been invested in determining 
the deviation of the g-factor from two, see \cite {Beringer:1900zz}
for a recent review.  The extraordinary precision naturally imposes constraints on
extensions of the Standard Model (SM), whose contribution $\Delta a_\mu$ to 
the anomalous magnetic moment 
\begin{align}
a_\mu=\frac{g_\mu-2}{2}=a^{\rm SM}_\mu+\Delta a_\mu
\end{align}
should not increase the observed slight tension between 
SM and experimental value for $a_\mu$.  

Among the more intensively studied classes of models beyond 
the SM are those that extend the dimension of space-time.   
A particularly rich phenomenology is realised in models that 
live on a compact slice of a five-dimensional Anti-de--Sitter space.  
In conformal coordinates the invariant interval of this space-time is given by 
\begin{align}
ds^2= \left(\frac{1}{kz}\right)^2 \left( \eta_{\mu\nu} dx^\mu dx^\nu-dz^2\right)\,,
\end{align}
where $k$ is of the order of the Planck mass $M_{\rm Pl}$ and the finite fifth dimension
stretches from $z=1/k$ to $z=1/T$. The a priori arbitrary scale $T$ is chosen 
to be of the order of a TeV to alleviate the gauge-gravity hierarchy problem \cite{Randall:1999ee}. 
While the original Randall-Sundrum (RS) set-up \cite{Randall:1999vf}
considered an extra-dimension that is only accessible to gravity, making the 
five-dimensional bulk accessible to SM fields 
\cite{Davoudiasl:1999tf,Pomarol:1999ad,Grossman:1999ra,Chang:1999nh} 
offers a geometric interpretation of the hierarchies for masses and flavour mixing angles 
in the fermion sector \cite{Pomarol:1999ad,Chang:1999nh,Huber:2000ie}---only 
the Higgs needs to be confined near $z=1/T$ (IR brane).

  The most immediate consequences of a compact extra-dimension is the 
  presence of higher excited modes for each SM field allowed to propagate in the bulk of AdS$_5$.
  The mass scale of these Kaluza-Klein (KK) resonances is set roughly by $2.5 \times T$. Direct production
  currently only yields lower limits on $T$ of the order of $\approx 1\, \rm TeV$ \cite{LHClimits}.
  Indirect constraints
  from flavour and electroweak precision observables give far stronger bounds. To avoid, in particular,  
  large corrections to the $\rho$ parameter and the $Zbb$ vertex one can extent the gauge group  
  in the bulk to SU(3)$_{c}\times$SU(2)$_{L}\times$SU(2)$_{R}\times$U(1)$_{X}\times$ $\mathds{Z}_2$
  with appropriate breaking  mechanisms on the branes to introduce a custodial protection into the model 
 \cite{Agashe:2003zs,Agashe:2006at,Albrecht:2009xr,Casagrande:2010si}.
  
  In the last few years, there have been increasing efforts to study the effects of the RS model  
  not only in processes that are dominated by tree-level effects, but also
  via observables that are sensitive to loop-induced processes.  
  Here, processes that are not expected to exhibit a UV sensitivity have been most promising; 
  notable studies include e.g.~Higgs production and decay 
  \cite{Hahn:2013nza,Malm:2013jia,Carena:2012fk,Azatov:2010pf},
  lepton \cite{Csaki:2010aj} and quark flavour violation 
  \cite{Gedalia:2009ws,Delaunay:2012cz,Blanke:2012tv}.
  The anomalous magnetic moment in the minimal RS model has been studied recently in 
  a 5D formalism in \cite{Beneke:2012ie}.  In this letter we will extend the results of \cite{Beneke:2012ie}
  to the custodially protected RS model. Here it should be noted that the model itself is 
  not UV complete. The effect of the unknown UV completion on the anomalous magnetic 
  moment is determined by the warped cut-off $\Lambda$ which is of the order of several times $T$.
  Generically one expects the correction to scale as ${1}/{\Lambda}$ and thus, provided that no
  additional enhancement factors are present, to be subleading.    
  
 This work is organised as follows. In Section 2 we describe the custodial protected RS model 
 and give the properties of the various bulk fields.  Our strategy for the calculation of $(g-2)_\mu$
 via matching the 5D theory onto an effective 4D lagrangian is detailed in Sect.~3. We follow \cite{Beneke:2012ie}
 and determine the Wilson  coefficients of the relevant SU(2)$\times$U(1) invariant dimension-six operators
 in a manifestly five-dimensional formalism.
 The various contributions are classified according to their dependence on the Yukawa structure of the 5D theory.
Our numerical results for the anomalous moment are presented in Section 4. We discuss their 
dependence on the parameters of the 5D Lagrangian and the impact of corrections to other 
lepton observables. We conclude in Section 5.

\section{The custodially protected model}

The standard way to implement the custodial protection
mechanism in the RS model is by extending gauge group in the bulk of AdS to
$$SU\left(3\right)_{c}\times SU\left(2\right)_{L}\times SU\left(2\right)_{R}\times U\left(1\right)_{X}\times \mathds{Z}_2 $$
where the discrete $\mathds{Z}_2$ enforces a symmetry under $SU\left(2\right)_{L} \leftrightarrow SU\left(2\right)_{R}$.
This assures that the gauge coupling constants for both SU(2)  group are identical, $g_{L}=g_{R}=g$. 
In order to obtain the Standard Model gauge group and particle content 
in the low energy limit, the bulk symmetry must be broken on both branes. On the UV brane the breaking proceeds 
by boundary conditions, on the IR brane by via the  Higgs vacuum expectation value (vev).
The presence of the $SU(2)_R$ group prohibits large correction to the $\rho$ parameter, while the
the discrete symmetry protects the $Zbb$ vertex from large corrections due to the (IR localised) top,
provided that the quarks are arranged in certain multiplets \cite{Agashe:2006at}.

There is, in principle, no need to mirror this construction in the lepton sector as the corrections to 
the $Z\tau\tau$ vertex are generally mild. One of the more economic constructions would  
put the lepton doublets in a $(2,1)$ multiplet under $SU\left(2\right)_{L}\times SU\left(2\right)_{R}$
and the singlets in a $(1,2)$ \cite{Carena:2007ua}. This realisation does not provide any genuinely 
new diagram topologies compared to minimal RS model and no significantly
enhanced contribution to $g_\mu-2$ can be expected\footnote{A simple estimate shows that the gauge contribution to 
$g_\mu-2$ is enhanced only by a factor of about $25\%$.}. We, therefore, turn to a model 
where the lepton sector is a (colourless) copy  of the quark sector \cite{Ledroit:2007ik}.

A detailed discussion of the model can be found in \cite{Albrecht:2009xr,Duling}. For our purposes 
it is enough to specify the action and the boundary conditions for the various 
fields on the branes; we follow the conventions of \cite{Duling}.  The action of the 
custodially protected model is given by
\begin{align}\label{eq_action}
 S_{(5D)}= &\int d^4x \int_{1/k}^{1/T} dz \sqrt{G} 
            \left \lbrace -\frac{1}{4} X^{MN} X_{MN} -\frac{1}{4} W_{R}^{a,MN} W^a_{R,MN}  -\frac{1}{4}  W_{L}^{a,MN}  W^a_{L,MN}  
               \right.     \nn \\
          & \left.+\sum_{\psi=\xi_1,\xi_2,T_4,T_3}  \left( e^M_m \left[ \frac{i}{2} \bar \psi^f \Gamma^m
                                  (\overrightarrow{D}_M - \overleftarrow{D}_M)\psi^f- c_{\psi_f} k   \bar \psi_f \psi_f \right] \right) \right \rbrace +S_{GF+ghost}      
                                                                                        \nn \\
     & +  \int  d^4x   \left \lbrace   (D^{\mu} \Phi)^\dagger D_\mu\Phi -V(\Phi)                             -\left(  \frac{T^3}{k^3} \right)\mathcal{L}_{\rm Yukawa}\right\rbrace            
\end{align}
where $G=1/(kz)^{10}$ and  $e^M_m=kz\, \text{diag}(1,1,1,1,1)$. The Higgs potential is given by 
\begin{align}
V(\Phi) = - \mu^2_{\rm (5D)}  \left(\frac{T}{k}\right)^{\!2} 
\Phi^\dagger\Phi+\frac{\lambda}{4}  \,(\Phi^\dagger\Phi)^2\,.
\end{align}
The covariant derivative is defined as
\begin{align}
D_{M}& =  \partial_{M}-ig_{X}Q_{X}X_{M} -ig_{5} T_L^{c}W_{L\, M}^{c}-ig_{5} T_R^{c} W_{R\, M}^{c}\,.
\label{eq_covariantDerivative}
\end{align}
$Q_X$ is the charge under the U(1)$_X$ whereas $T^c_{R/L}$ are the generators of the SU(2)$_{L/R}$
in the appropriate representation: $\xi_1$ is a bi-doublet under  SU(2)$_{L}\times$ SU(2)$_{R}$, $\xi_2$ 
is a singlet and $T_{3/4}$ are singlets under SU(2)$_{L}$ and triplets under  SU(2)$_{R}$. 
The lepton multiplets can conveniently be written in the form \cite{Albrecht:2009xr} 
\begin{eqnarray}
\xi_{1L}^{il} & = & \left(\begin{array}{cc}
\chi_{L}^{\nu_{i}}\left(-,+\right)_{1} & l_{L}^{\nu_{i}}\left(+,+\right)_{0}\\
\chi_{L}^{l_{i}}\left(-,+\right)_{0} & l_{L}^{l_{i}}\left(+,+\right)_{-1}
\end{array}\right)\nonumber \\
\xi_{2R}^{il} & = & \nu_{R}\left(+,+\right)_{0}\nonumber \\
\xi_{3R}^{il} & = & T_{3R}^{i}\otimes T_{4R}^{i}=\left(\begin{array}{c}
\tilde{\lambda}_{R}^{_{i}}\left(-,+\right)_{1}\\
\tilde{N}_{R}^{_{i}}\left(-,+\right)_{0}\\
\tilde{L}_{R}^{_{i}}\left(-,+\right)_{-1}
\end{array}\right)\otimes\left(\begin{array}{c}
\lambda_{R}^{_{i}}\left(-,+\right)_{1}\\
N_{R}^{_{i}}\left(-,+\right)_{0}\\
E_{R}^{_{i}}\left(+,+\right)_{-1}
\end{array}\right),\label{eq:T2}
\end{eqnarray}
where the subscript of the different fields indicates the electrical
charge $Q =T_{L}^{3}+T_{R}^{3}+Q_{X}$. The signs in parentheses 
are the boundary conditions on the UV (left) and IR brane (right); 
a '+' corresponds to Neumann and a '-' to Dirichlet boundary conditions.
Only $(+,+)$ fields have massless zero-modes. The Standard model doublet field is contained in  $\xi_{1}$, while the 
singlet field is included in the $T^{4}$ triplet.
Note that this model additionally includes a right--handed neutrino. The field strength tensors are given by the usual expressions 
\begin{align}
X_{{NM}}=\partial_{N} X_{M}-\partial_{M} X_{N} && W_{R/L, NM}^a=\partial_{N} W_{R/L,M}^a-\partial_{M} W_{R/L,N}^a +g_5\varepsilon^{abc} W_{R/L,N}^bW_{R/L,M}^c\;. 
\end{align} 
The SM $B$ boson arises as the combination
\begin{eqnarray}
W_{R}^{3} & = & \cos\phi\, Z_{X}+\sin\phi B\nonumber \\
X & = & -\sin\phi\, Z_{X}+\cos\phi\, B \label{eq:A5}
\end{eqnarray}
in analogy to the $W_L^3$-$B$ mixing in the SM. In particular 
it follows that 
\begin{eqnarray}
\cos\phi=\frac{g_{5}}{\sqrt{g_{5}^{2}+g_{5X}^{2}}} 
\end{eqnarray}
and 
\begin{equation}
g'_{5}=\frac{g_{5}\, g_{5X}}{\sqrt{g_{5}^{2}+g_{5X}^{2}}}.
\end{equation}
for the U(1) hypercharge coupling $g^\prime_5$. The corresponding 
boundary conditions for the four-vector components of the gauge fields are  
\begin{eqnarray}
W_L^{\mu,a}\left(+,+\right) & , & W_{R}^{\mu,b}\left(-,+\right)\\
B^{\mu}\left(+,+\right) & , & Z_{X}^{\mu}\left(-,+\right)
\end{eqnarray}
with $a=1,2,3$ and $b=1,2$.

The Yukawa interaction is described by the Lagrangian density \cite{Albrecht:2009xr}
\begin{align}
\mathcal{L}_{\rm Yukawa} =& \sum_{i,j=1}^{3}\left \lbrace -(\lambda_{5d}^{u})_{ij}\,\left(\bar{\xi}_{1}^{i}\right)_{a\alpha}\Phi_{a\alpha}\xi_{2}^{j} \phantom{\frac{T}{k}}\right.\nonumber \\
 & \left. +\sqrt{2}\,\lambda_{5d\, ij}^{d}\left[\left(\bar{\xi}_{1}^{i}\right)_{a\alpha}\frac{\left(\tau^{c}\right)_{ab}}{2}\,\left(\tilde{T}_{3}^{j}\right)_{c}\Phi_{b\alpha}+\left(\bar{\xi}_{1}^{i}\right)_{a\alpha}\frac{\left(\tau^{\gamma}\right)_{\alpha\beta}}{2}\,\left(\tilde{T}_{4}^{j}\right)_{\gamma}\Phi_{a\beta}\right]+\text{h.c.}\right\rbrace
\end{align}
where we show all group indices explicitly. The $\tilde{T}_{3,4}$ are defined as

\begin{eqnarray}
\left(\tilde{T}_{3}^{i}\right)=\left(\begin{array}{c}
\frac{1}{\sqrt{2}}\left(\tilde{\lambda}^{_{i}}+\tilde{L}^{_{i}}\right)\\
\frac{i}{\sqrt{2}}\left(\tilde{\lambda}^{_{i}}-\tilde{L}^{_{i}}\right)\\
\tilde{N}^{_{i}}
\end{array}\right)\qquad\left(\tilde{T}_{4}^{i}\right)=\left(\begin{array}{c}
\frac{1}{\sqrt{2}}\left(\lambda^{_{i}}+E^{_{i}}\right)\\
\frac{i}{\sqrt{2}}\left(\lambda^{_{i}}-E^{_{i}}\right)\\
N^{i}
\end{array}\right)
\end{eqnarray} in the basis of (\ref{eq:T2}).
Note that the Higgs is not charged under the U(1)$_X$  but is
a bi-doublet under the SU(2) groups:
\begin{equation}
\Phi=\left(\begin{array}{cc}
\pi^{+}/\sqrt{2} & -\left(v+h^{0}-i\pi^{0}\right)/\sqrt{2}\\
\left(v+h^{0}+i\pi^{0}\right)/\sqrt{2} & \pi^{-}/\sqrt{2}
\end{array}\right),
\end{equation}
where $h^0=\pi^\pm=0$ gives the vacuum expectation value of the bi-doublet. 

In writing \eqref{eq_action} we effectively assumed that the Higgs field 
is delta-function localised on the IR brane
and eliminated the integral over the bulk coordinate $z$ using the $\delta(g_{55} (z-1/T))$ factor. 
As has been pointed out numerously in the literature
\cite{Azatov:2010pf,Delaunay:2012cz,Carena:2012fk} the localisation should be
implemented via some limiting procedure.
 To completely specify the model one needs to give a prescription 
 how to take the limit of the regulator for the $\delta$-distribution. 
We will discuss this in section \ref{sec:higgs}.


\section{Matching  onto an effective  dimension six lagrangian}

We will follow the multi-step matching strategy prescribed in \cite{Beneke:2012ie},
 i.e.~we integrate out the effect of the heavy KK fields and match onto a 
manifestly SU$(3)_c \times$SU$(2)_L\times$U$(1)$ symmetric Lagrangian. Hence, 
we will be brief and refer the reader to  \cite{Beneke:2012ie} for a detailed discussion.
The dominant effects beyond the Standard Model can then be captured 
by operators of dimension six\footnote{The single dimension-five operator is not
relevant for our analysis.}  \cite{Buchmuller:1985jz,  Grzadkowski:2010es}.  
\begin{align}
 \mathcal{L}^{RS_{cs}}\to \mathcal{L}^{SM} +\frac{1}{T^2}\sum_i c_i {\cal O}_i\,
\end{align}
where 
\begin{align}\label{dimsixlagragian}
 \sum_i c_i {\cal O}_i=   
   & \phantom{+} a_B^{ij} \bar L_i \sigma^{\mu \nu} E_j  \Phi B_{\mu\nu} 
       + a_W^{ij} \bar L_i \tau^A \sigma^{\mu \nu} E_j  \Phi W^A_{\mu\nu}  + \text{h.c.} 
\nn \\
   &   + b^{ij} \bar L_i\gamma^\mu L_i \; \bar E_j \gamma_\mu E_j  
\nn \\ 
   &    
         + c_1^{i} \; \Phi^\dagger i{D}_\mu \Phi   \; \bar E_i \gamma^\mu E_i 
         + c_2^{i} \; \Phi^\dagger i{D}_\mu \Phi   \; \bar L_i \gamma^\mu L_i 
         + c_3^{i} \; \Phi^\dagger i\overleftrightarrow{\tau^A D_\mu} \Phi \; \bar L_i \tau^A \gamma^\mu L_i 
         \nn \\
   &   + h^{ij}  \Phi^\dagger \Phi\; \bar L_i \Phi E_j   + \text{h.c.}
\end{align}
where $\overleftrightarrow{\tau^A D_\mu}=1/2(\tau^A \overrightarrow{D}_\mu- \overleftarrow{D}_\mu \tau^A)$.
$L_i$ represents a  lepton doublet field of flavour $i$ and 
$E_i$ stands for the SM lepton singlet. The Higgs doublet is given by $\Phi$ 
and $ B_{\mu\nu}$ and $W^A_{\mu\nu}$ are the usual field strength tensors of 
 U(1)$_{\textrm{Y}}$ and  SU(2)$_{\textrm{L} }$ gauge field, respectively. 
The covariant derivative is given by
\begin{align}
D_\mu=\partial_\mu-i g^\prime \frac{Y}{2} B_\mu-ig_5 T^a W^a_\mu
\end{align}
with $Y$ being the hypercharge operator and $T^a$ being the generators of SU(2)
in the appropriate representation.  
\eqref{dimsixlagragian} contains  only those operators that
 can either contribute to the anomalous magnetic moment at the one loop-level
in the effective theory  and can be generated at tree level in the full theory 
(last three lines), or they 
contribute at tree-level but are generated by loops in the 5D theory (first line).
 
The basic formula \cite{Beneke:2012ie}, equation (2.26) therein, for the
 anomalous magnetic moment is 
\begin{align}
\Delta a_\mu 
&= -\,\frac{4 m_\mu^2}{T^2}\left(
\frac{\mbox{Re}\,(\alpha_{22})}{y_\mu e}
+\sum_{k=1,2,3} \frac{1}{16\pi^2}\,
\frac{m_{\ell_k}}{m_\mu}\,
\mbox{Re}\,(\beta_{2kk2})
\right)
\label{amures}
\end{align} 
 with 
 \begin{align}
\alpha_{ij} &= [U^\dagger a V]_{ij},
\nonumber\\[0.1cm]
a_{ij}&=\cos\Theta_W a^{B,ij}-\sin\Theta_W a^{W,ij},
\nonumber\\[0.1cm]
\beta_{ijkl} &= \sum_{m,n} \,
[U^\dagger]_{im} U_{mj} [V^\dagger]_{kn} V_{nl} \,b^{mn} ,
\label{couplingdefs}
 \end{align}
where $U$, $V$ are the rotation matrices into the mass eigenbasis for 
charged doublet and singlet leptons, respectively.
The contribution proportional to the Wilson coefficient $h$  is not shown 
as it is suppressed by a factor of $m_\ell/v$.
\eqref{amures} does not include effects that can directly or indirectly be 
associated with modifications
of SM parameters. 
\begin{figure}
\begin{center}
 \includegraphics[width=9cm]{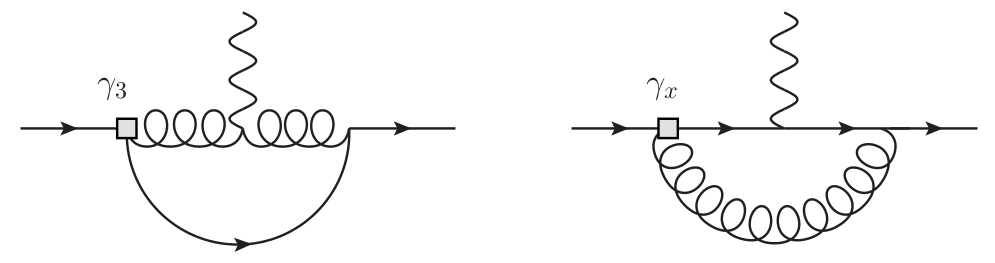}
\end{center}
 \caption{\label{fig:Zcouplings} SM electroweak diagrams 
(in unitary gauge) with insertions of 
operators of the form $\bar \psi \gamma^\mu \psi \cdot \Phi^\dagger D_\mu \Phi$. 
Diagrams with an operator insertion in the other vertex are not shown.}
\end{figure}
In particular, the direct modification of the Z-muon and W-muon 
couplings\footnote{Assuming one does not use e.g.~the 
measured values for the W/Z-muon couplings as input parameters.} via the 
Wilson coefficients $c_{1,2,3}^i$ was omitted. They enter via the diagrams
 shown in Fig.~\ref{fig:Zcouplings} which arise from
insertions of dimensions-six operators in SM one-loop diagrams. Their effect
 on $g-2$ is given by 
\begin{align}
 \Delta a_\mu^{ZW}&=  -\,\frac{4 m_\mu^2}{T^2}  \frac{1}{3(4\pi)^2} \; {\rm Re}\left[ \gamma_2^{22}- \frac32\gamma^{22}_1 -\frac32 \gamma^{22}_3
 +\sin^2\Theta_W \sum_{i=1}^3 \gamma_i^{22} \right]
\end{align}
with \begin{align}
      \gamma_1^{ij}=\sum_k[V^\dagger]_{ik} c^k_1 V_{kj}&& \gamma_x^{ij}=\sum_k[U^\dagger]_{ik} c^k_x U_{kj}, \;x=2,3\,,
     \end{align}
see also \cite{Crivellin:2013hpa} where different conventions for operator normalisation, 
covariant derivative and momentum flow are used.
Furthermore, there are indirect effects that stem from modifications of SM relations. 
For example, the SM electroweak corrections to $a_\mu$ are usually written in terms of $G_F$.
However, the usual relation of $W$ mass, electroweak coupling $g$ and $G_F$ receives
corrections from dimension-six operators, see e.g.~\cite{Alonso:2013hga}. These indirect effects
depend on the choice of input parameters and, while readily calculable, are omitted here. 

We now can relate the Wilson coefficients in \eqref{dimsixlagragian} to the shift in  
$(g-2)_\mu$. This leaves us with the determination of the Wilson coefficients in the (custodial) RS model.   

\subsection{Gauge Contributions}

The terms contributing to  the Wilson coefficients in the dimension--six Lagrangian that are at most linear in the 
Yukawa couplings necessarily arise from diagrams that contain at least one exchange of a 5D gauge boson. 
We therefore usually refer to these terms as {\it gauge contributions}. On the other hand 
there are terms that involve three Yukawa factors; these
are usually generated by an exchange of a Higgs boson and hence dubbed {\it Higgs contributions}.
We will discuss the two separately.
The gauge contributions in the minimal model were discussed at length in \cite{Beneke:2012ie}. Most of these results 
carry over to the custodially protected model. We only need to account for effects that originate from 
the additional particles in the spectrum.

\begin{figure}
 \begin{center}
 \includegraphics[width=0.25\textwidth]{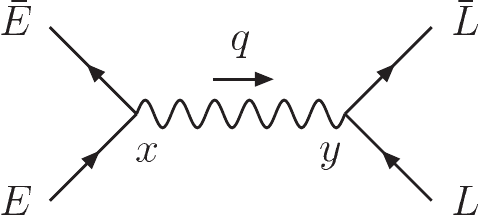} \hspace{1cm} 
 \includegraphics[width=0.30\textwidth]{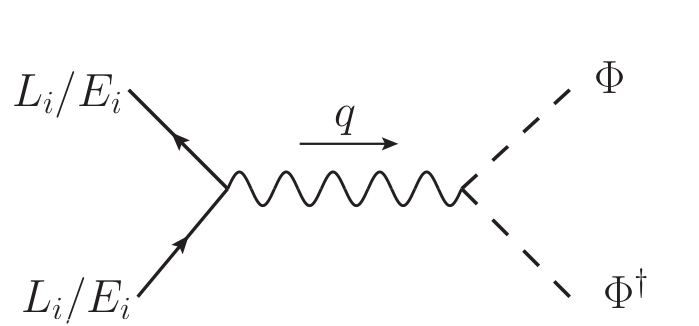}
\end{center}
 \caption{Additional diagrams for the tree-level matching calculation. 
          The internal gauge boson is a $Z_X$.\label{fig_fourfermion}}
\end{figure}

{In the case of operators that can be 
generated at tree-level in the 5D theory} there are only three additional diagrams that influence the 
matching calculation. The reason for this is that 
an interaction with the new non-abelian gauge bosons $W_R^{1,2}$ always changes the 
SU(2)$_R$ quantum number of leptons in such a way that at least one of the 
fermions must not have a zero-mode. Only the $Z_X$ boson can modify the Wilson coefficients 
of tree-level operators relative to their value in the minimal model. 
The diagrams that need to be evaluated are shown in figure \ref{fig_fourfermion}. 
Note, that the fifth component of the $Z_X$
cannot appear as the external modes at each vertex have the same handedness.

Since the external momenta are always much smaller than the KK scale $T$ we only need the 
expression for the $Z_X^{\mu}$ propagator in the limit of vanishing 4D momentum $q$:
\begin{align}
 \Delta_\perp^{Z_X}(q\to 0,x,y)= \frac{i}{2k}\left[ k^2\,\text{min}\!\left\lbrace   x^2,y^2  \right\rbrace -1 \right ]\;.
\end{align}
This expression can be obtained from \cite{Beneke:2012ie} by taking into account the 
modified boundary condition on the UV brane.
In the following we give the complete expressions for the Wilson coefficients $b_{ij}$ and $c_x$, $x=1,2,3$.
We dropped terms that are suppressed by powers of the tiny ratio $\epsilon=T/k$.
The first line always  gives the contribution due to $B$ and $W_L$ boson exchanges that are already 
present in the minimal model; the second line, if present, represents the contribution 
from the new bosons.
\begin{align}
  c_{1,i}= &\;\frac{{g^\prime}^2 Y_E}{8} \,\Bigg(1-\frac{1}{\ln1/\epsilon}  
    - \left[ \frac{(1+2c_{E_i})(5+2c_{E_i})}{(3+2c_{E_i})^2} 
    -\frac{2(1+2c_{E_i})\ln1/\epsilon}{(3+2c_{E_i})} \right]
    \frac{\epsilon^{-2c_{E_i}-1}}{1-\epsilon^{-2c_{E_i}-1}}\Bigg) \nn \\
     &+ \frac{(g^2-{g^\prime}^2)Y_E }{4}  \left[ \frac{(1+2c_{E_i})\ln1/\epsilon}{(3+2c_{E_i})}\right]
        \frac{\epsilon^{-2c_{E_i}-1}}{1-\epsilon^{-2c_{E_i}-1}}
\label{c1coeff}
\end{align}
\begin{align}
  c_{2,i}=  &\;\frac{{g^\prime}^2 Y_L}{8} \,\Bigg(1-\frac{1}{\ln1/\epsilon}  
    - \left[ \frac{(1-2c_{L_i})(5-2c_{L_i})}{(3-2c_{L_i})^2} 
    -\frac{2(1-2c_{L_i})\ln1/\epsilon}{(3-2c_{L_i})} \right]
    \frac{\epsilon^{2c_{L_i}-1}}{1-\epsilon^{2c_{L_i}-1}}\Bigg) \nn \\
    & + \frac{(g^2-{g^\prime}^2)Y_L}{4}  \left[ \frac{(1-2c_{L_i})\ln1/\epsilon}{ 3-2c_{L_i} }\right]
        \frac{\epsilon^{2c_{L_i}-1}}{1-\epsilon^{2c_{L_i}-1}}
\label{c2coeff}
\end{align}
and
\begin{align}
  c_{3,i}= \;\frac{g^2}{8} \,\Bigg(1-\frac{1}{\ln1/\epsilon}  
    - \left[ \frac{(1-2c_{L_i})(5-2c_{L_i})}{(3-2c_{L_i})^2} 
    -\frac{2(1-2c_{L_i})\ln1/\epsilon }{ 3-2c_{L_i} } \right]
    \frac{\epsilon^{2c_{L_i}-1} }{1-\epsilon^{2c_{L_i}-1} }
\Bigg).
\label{c3coeff}
\end{align}
$Y_E$ and $Y_L$ are the hypercharges of singlet and doublet, respectively. 
The coefficient for the four-lepton operator $\bar L_i \gamma_\mu L_i \cdot \bar E_j \gamma^\mu E_j$ is given by
\begin{align}     
b_{ij} = &\; b_0 + b_1(c_{L_i}) + b_1(-c_{E_j}) + b_2(c_{L_i},c_{E_j}) \nn \\
         & +  \frac{g^2-{g^\prime}^2}{{g^\prime}^2} b_2(c_{L_i},c_{E_j})
\end{align}
with
\begin{align}
&b_0 = - \frac{ {g^{\prime}}^2}{8}\,\frac{1}{\ln(1/\epsilon)} ,
\nonumber\\
& b_1(c) = - \frac{ {g^{\prime}}^2}{8}\,\frac{(5-2 c)(1-2 c)}{(3-2
  c)^2} \,\frac{\epsilon^{2 c-1}}{1-\epsilon^{2 c-1}},\nn
\\
 & b_2(c_L,c_E) = 
 - \frac{ {g^{\prime}}^2}{4}\,
\frac{(1-2 c_L) (1+2 c_E)(3-c_L+c_E)}
{(3-2 c_L) (3+2 c_E)(2-c_L+c_E)}
 \,\ln\frac{1}{\epsilon}\,
 \frac{\epsilon^{2 c_L-1}}{1-\epsilon^{2 c_L-1}}
 \,\frac{\epsilon^{-2 c_E-1}}{1-\epsilon^{-2 c_E-1}}.
\end{align}
Note that for UV localised lepton zero-modes the Wilson coefficients
$c_i$ are dominated by the '$1$' in the first line. All flavour dependent terms 
are then subleading.  

\begin{figure}
\begin{minipage}{0.03\textwidth}
{\footnotesize \mbox{ }\hspace{0.25cm} }\\
\mbox{ }
\end{minipage}
\begin{minipage}{0.205\textwidth}\center{A1}\\
\includegraphics[width=1.0\textwidth]{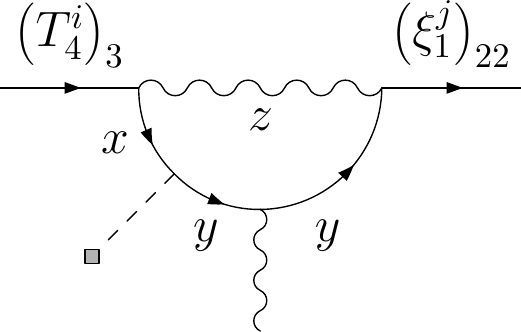}
\end{minipage}
\begin{minipage}{0.03\textwidth}
{\footnotesize \mbox{ }\hspace{0.25cm} }\\
\mbox{ }
\end{minipage}
\begin{minipage}{0.205\textwidth}\center{A2}\\
\includegraphics[width=1.0\textwidth]{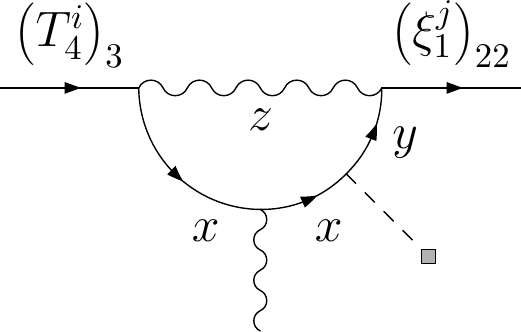}
\end{minipage}
\begin{minipage}{0.03\textwidth}
{\footnotesize \hspace{0.25cm} }\\
\mbox{ }
\end{minipage}
\begin{minipage}{0.205\textwidth}\center{A3}\\
\includegraphics[width=1.0\textwidth]{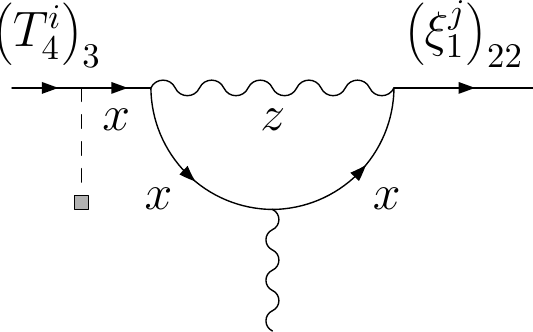}
\end{minipage}
\begin{minipage}{0.03\textwidth}
{\footnotesize \hspace{0.25cm} }\\
\mbox{ }
\end{minipage}
\begin{minipage}{0.205\textwidth}\center{A4}\\
\includegraphics[width=1.0\textwidth]{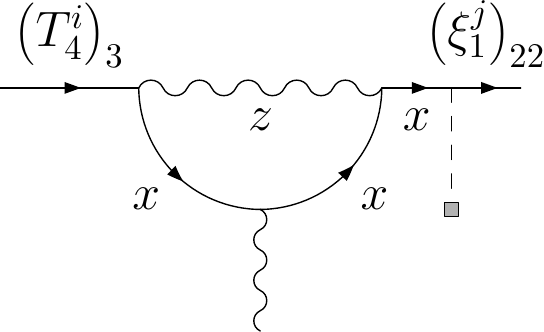}
\end{minipage}
\newline
\mbox{ }
\newline
\mbox{ }
\newline
\begin{minipage}{0.03\textwidth}
{\footnotesize \hspace{0.25cm} }\\
\mbox{ }
\end{minipage}
\begin{minipage}{0.205\textwidth}\center{A5}\vspace{0.2cm}\\
\includegraphics[width=1.0\textwidth]{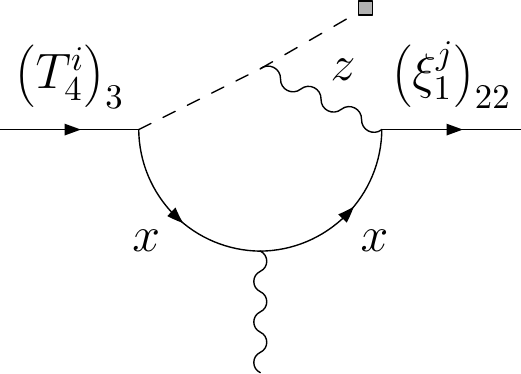}
\end{minipage}
\begin{minipage}{0.03\textwidth}
{\footnotesize \hspace{0.25cm} }\\
\mbox{ }
\end{minipage}
\begin{minipage}{0.205\textwidth}\center{A6}\vspace{0.2cm}\\
\includegraphics[width=1.0\textwidth]{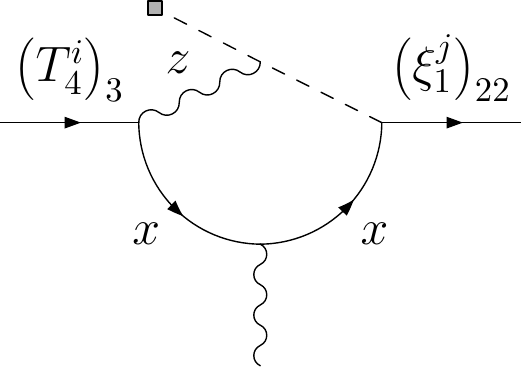}
\end{minipage}
\begin{minipage}{0.05\textwidth}

\end{minipage}
\begin{minipage}{0.205\textwidth}

\end{minipage}
\begin{minipage}{0.05\textwidth}

\end{minipage}
\begin{minipage}{0.205\textwidth}

\end{minipage}
\newline
\mbox{ }
\newline
\mbox{ }
\newline
\mbox{ }
\newline
\begin{minipage}{0.03\textwidth}
{\footnotesize \hspace{0.25cm} }\\
\mbox{ }
\end{minipage}
\begin{minipage}{0.205\textwidth}\center{W1}\vspace{0.2cm}\\
\includegraphics[width=1.0\textwidth]{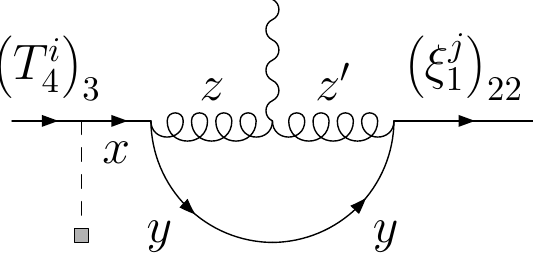}
\end{minipage}
\begin{minipage}{0.03\textwidth}
{\footnotesize \hspace{0.25cm} }\\
\mbox{ }
\end{minipage}
\begin{minipage}{0.205\textwidth}\center{W2}\vspace{0.2cm}\\
\includegraphics[width=1.0\textwidth]{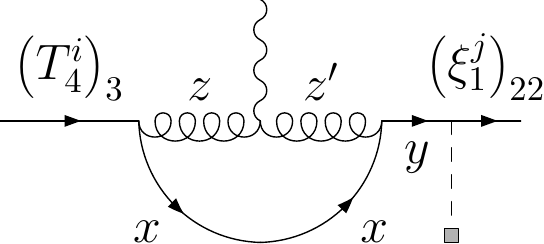}
\end{minipage}
\begin{minipage}{0.03\textwidth}
{\footnotesize \hspace{0.25cm} }\\
\mbox{ }
\end{minipage}
\begin{minipage}{0.205\textwidth}\center{W3}\vspace{0.2cm}\\
\includegraphics[width=1.0\textwidth]{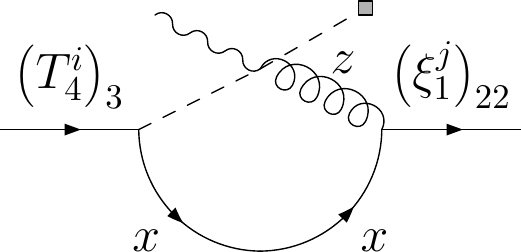}
\end{minipage}
\begin{minipage}{0.03\textwidth}
{\footnotesize \hspace{0.25cm} }
\\
\mbox{ }
\end{minipage}
\begin{minipage}{0.205\textwidth}\center{W4}\vspace{0.2cm}\\
\includegraphics[width=1.0\textwidth]{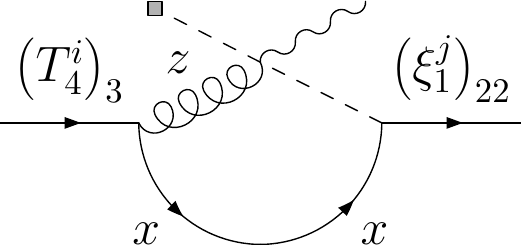}
\end{minipage}
\mbox{ } 
\newline
\mbox{ }
\newline
\mbox{ }
\newline
\mbox{ }
\newline
\begin{minipage}{0.03\textwidth}
{\footnotesize \hspace{0.25cm} }\\
\mbox{ }
\end{minipage}
\begin{minipage}{0.205\textwidth}\center{W5}\vspace{0.2cm}\\
\includegraphics[width=1.0\textwidth]{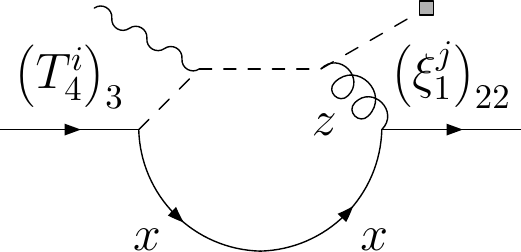}
\end{minipage}
\begin{minipage}{0.03\textwidth}
{\footnotesize \hspace{0.25cm} }\\
\mbox{ }
\end{minipage}
\begin{minipage}{0.205\textwidth}\center{W6}\vspace{0.2cm}\\
\includegraphics[width=1.0\textwidth]{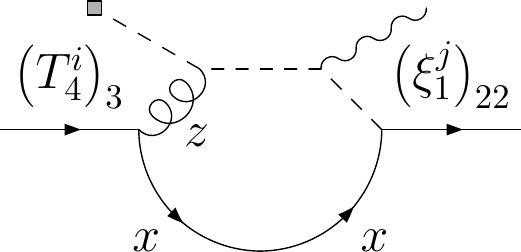}
\end{minipage}
\begin{minipage}{0.03\textwidth}
{\footnotesize \hspace{0.25cm} }\\
\mbox{ }
\end{minipage}
\begin{minipage}{0.205\textwidth}\center{W7}\vspace{0.2cm}\\
\includegraphics[width=1.0\textwidth]{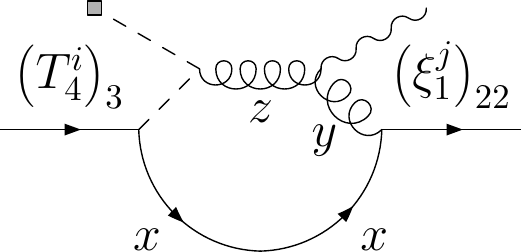}
\end{minipage}
\begin{minipage}{0.03\textwidth}
{\footnotesize \hspace{0.25cm} }
\\
\mbox{ }
\end{minipage}
\begin{minipage}{0.205\textwidth}\center{W8}\vspace{0.2cm}\\
\includegraphics[width=1.0\textwidth]{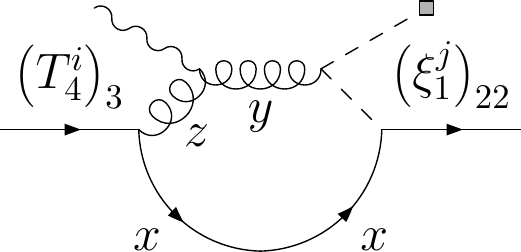}
\end{minipage}
\mbox{ } 
\newline
\mbox{ }
\newline
\mbox{ }\vspace{0.2cm}
\newline
\begin{minipage}{0.03\textwidth}
{\footnotesize \hspace{0.25cm} }\\
\mbox{ }
\end{minipage}
\begin{minipage}{0.205\textwidth}\center{W9}\vspace{0.2cm}\\
\includegraphics[width=1.0\textwidth]{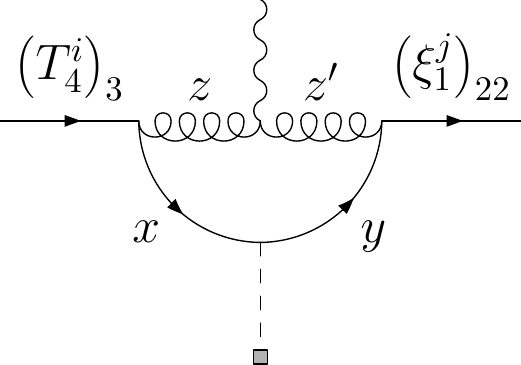}
\end{minipage}
\begin{minipage}{0.03\textwidth}

\end{minipage}
\begin{minipage}{0.205\textwidth}

\end{minipage}
\begin{minipage}{0.03\textwidth}

\end{minipage}
\begin{minipage}{0.205\textwidth}

\end{minipage}
\begin{minipage}{0.03\textwidth}

\end{minipage}
\begin{minipage}{0.205\textwidth}

\end{minipage}
\caption{\label{fig_collectedtopologies} All topologies with internal gauge bosons that contribute to the matching of
the dipole operator Wilson coefficient at one loop. Abelian topologies are labelled by A1-6, non-abelian
topologies by W1-10. Fermions represented by straight lines and Higgs bosons by dashed lines.  The final and initial 
fermions are always assumed to be $(T_4)_3$ and $(\xi_1)_{22}$---the fields corresponding to the SM singlet and doublet.
$x,y,z,z^\prime$ label the species of the internal propagators, see tables \ref{tab:Possible-field-configurationA} and \ref{tab:Possible-field-configurationNA} for the the allowed particles.}
\end{figure}

{The Wilson coefficients for the dipole operators } are more 
complicated to obtain than the Wilson coefficients
generated by tree-level exchanges. We need to compute 5D one-loop diagrams.
The diagram topologies  are shown in figure \ref{fig_collectedtopologies}, 
while the allowed assignments for the particles\footnote{Note that $\xi_2$ cannot propagate. 
Hence, only the Yukawa $\lambda^d$ can appear in the Wilson coefficient.} in each topology are given 
in tables \ref{tab:Possible-field-configurationA} and \ref{tab:Possible-field-configurationNA}.
Our strategy for evaluating the 5D loops is identical to the one given 
in \cite{Beneke:2012ie}, section 3.3 therein, and we refer 
the reader to the reference for the technical details. 
In particular, the analytic scheme-independence and gauge-invariance proofs 
proceeds in full analogy to the calculation in the minimal model. 
Still some comment on the 5D gauge parameter $\xi$ is in order.
It had to be introduced to disentangle the fifth component of the 
gauge field from the vector components under
free propagation \cite{Randall:2001gb}. 
The Wilson coefficients of the dipole operators have to be independent of 
the choice for the gauge fixing term and therefore of $\xi$. Since we can choose different 
gauge parameters for, e.g.,~abelian and non-abelian fields 
we can use gauge-invariance as a separate check for the contribution
of individual gauge fields.
In all calculations we work in general $R_\xi$ gauge and keep the gauge parameter $\xi$
as a free variable. For a more detailed discussion of gauge-invariance and the importance of 
one-particle reducible diagrams see \cite{Beneke:2014sta}.

The main difference to the calculation in the minimal model is that the topology of a 
diagram is not enough to fix the structure of the integrals as we now have to deal with 
fermions and bosons  with different possible boundary conditions on the branes.
This substantially increases the computational effort of evaluating the Wilson coefficients, 
but does not lead to any additional conceptional difficulties compared to \cite{Beneke:2012ie}.

In \cite{Beneke:2012ie} the loop integrals, that is the integrals over the bulk positions
of each vertex as well as the integral over the modulus of the 4D loop momentum $l$
were carried out numerically. Given the large number of diagrams this strategy 
is no longer feasible in the custodially protected model. 
To improve our accuracy we use analytic solutions for all vertex integrals with external 
gauge bosons based on orthogonality and completeness relations. We then only have to
perform at most two bulk coordinate integrals and the $l$-integration numerically. 

A further decrease in the numerical uncertainty can be achieved by 
evaluating the remaining integrals in a two step procedure.
We first utilise routines provided by the CUBA library \cite{Hahn:2004fe} 
to perform the remaining coordinate integrals for fixed values of 
the modulus of the 4D loop-momentum. We choose a set of 2180 loop-momentum values.
These are used to construct an interpolation grid for the $l$-integrand. The first $2000$
points capture the structure of this integrand in the interval $l\in (0,10T)$.
In this region we cannot use simplifications and need to 
exclusively rely on the numerical results. For $l>10T$ we can already 
make use of the result for the integrand that is obtained if all
propagators are replaced by their asymptotic forms for $l\gg T$, i.e.,~products of hyperbolic 
sine and cosine functions. In this case the coordinate integrals can be 
solved analytically and the $l$ integrand has a simple form.  
We then use the remaining 180 points to map the deviations from the asymptotic form in the 
momentum region $10T<l<100T$. For $l>100T$ we can rely solely on the asymptotic expression.
The grid together with the asymptotic result are then used to construct an interpolating 
function of the integrand for all values of $l$. As a final step, this interpolating 
function is integrated with Mathematica.

This detour via an interpolating grid is superior to a simultaneous numerical integration of
all integration variables, i.e., the bulk coordinates and modulus of the
loop momentum. In general the difference between the two approaches is 
negligible. However, in diagrams with an exchange of a gauge boson that
has a zero-mode we observe subtle cancellations on the integrand level
that are only accurately resolved (for acceptable computer run-times)
in the grid approach. This issue only becomes numerically relevant for 
5D fermion mass parameters that force the corresponding fermion zero-modes 
to be localised towards the IR brane. Hence, the error introduced by the
naive integration approach is generally negligible for the study of 
lepton observables. However, the same integral structures appear, e.g., in
$b\to s\gamma$ and $t\to cg$ where the effect can be of the order of several percent.

In total the optimised integration routine is by more 
than a factor of $10$ faster and more accurate than the 'brute force' 
approach we used in \cite{Beneke:2012ie}.

\begin{table}
\begin{center}%
\begin{tabular}{|c|c|c|c||c|c|c|c|}
\hline 
 & x & y & z &  & x & y & z\tabularnewline
\hline 
\hline 
A1 & $\left(T_{4}^{i}\right)_{3}$ & $\left(\xi_{1}^{j}\right)_{22}$ & $B^{N}$ &A2& $\left(T_{4}^{i}\right)_{3}$ & $\left(\xi_{1}^{j}\right)_{22}$ & $B^{N}$\tabularnewline
\hline 
$\vphantom{\left(\xi_{1}^{j}\right)_{22}}$ & $\left(T_{4}^{i}\right)_{3}$ & $\left(\xi_{1}^{j}\right)_{22}$ & $Z_{X}^{N}$ &  & $\left(T_{4}^{i}\right)_{3}$ & $\left(\xi_{1}^{j}\right)_{22}$ & $Z_{X}^{N}$\tabularnewline
\hline 
\hline 
A3 & $\left(\xi_{1}^{j}\right)_{22}$ & / & $B^{N}$ & A4 & $\left(T_{4}^{i}\right)_{3}$ & / & $B^{N}$\tabularnewline
\hline 
 & $\left(\xi_{1}^{j}\right)_{22}$ & / & $Z_{X}^{N}$ & & $\left(T_{4}^{i}\right)_{3}$ & / & $Z_{X}^{N}$\tabularnewline
\hline 
& $\left(\xi_{1}^{j}\right)_{22}$ & / & $W_{L}^{3\, N}$ &  &  &  & \tabularnewline
\hline 
\hline 
A5 & $\left(\xi_{1}^{j}\right)_{22}$ & / & $B^{\mu}$ & A6& $\left(T_{4}^{i}\right)_{3}$ & / & $B^{\mu}$\tabularnewline
\hline 
 & $\left(\xi_{1}^{j}\right)_{22}$ & / & $Z_{X}^{\mu}$ &  & $\left(T_{4}^{i}\right)_{3}$ & / & $Z_{X}^{\mu}$\tabularnewline
\hline 
& $\left(\xi_{1}^{j}\right)_{22}$ & / & $W_{L}^{3\,\, \mu}$ &  &  &  & \tabularnewline
\hline 
\end{tabular}\end{center}
\caption{Possible field configuration inside the loop of the abelian diagram
topologies A1-A6 \label{tab:Possible-field-configurationA}. A capital roman index  on a gauge field indicates that 
both the vector and the scalar fifth component are valid options, a small Greek index shows that only the vector 
components may propagate.}
\end{table}

\begin{table}
\begin{center}%
\begin{tabular}{|c|c|c|c|c||c|c|c|c|c|}
\hline 
 & x & y & z & z$^{\prime}$ &  & x & y & z & z$^{\prime}$\tabularnewline
\hline 
\hline 
W1  & $\left(\xi_{1}^{j}\right)_{22}$ & $\left(\xi_{1}^{j}\right)_{12}$ & $W_{L}^{-\, N}$ & $W_{L}^{+\, N^{\prime}}$ &W2& $\left(T_{4}^{i}\right)_{2}$ & $\left(T_{4}^{i}\right)_{3}$ & $W_{R}^{-\, N}$ & $W_{R}^{+\, N^{\prime}}$\tabularnewline
\hline 
  & $\left(\xi_{1}^{j}\right)_{22}$ & $\left(\xi_{1}^{j}\right)_{21}$ & $W_{R}^{-\, N}$ & $W_{R}^{+\, N^{\prime}}$ &  &  &  &  & \tabularnewline
\hline 
W3 & $\left(\xi_{1}^{j}\right)_{12}$ & / & $W_{L}^{+\,\mu}$ & / & W4& $\left(T_{4}^{i}\right)_{2}$ & / & $W_{R}^{+\,\mu}$ & /\tabularnewline
\hline 
W5 & $\left(\xi_{1}^{j}\right)_{12}$ & / & $W_{L}^{+\,\mu}$ & / & W6 & $\left(T_{4}^{i}\right)_{2}$ & / & $W_{R}^{+\,\mu}$ & /\tabularnewline
\hline 
W7 & $\left(\xi_{1}^{j}\right)_{12}$ & $W_{L}^{+\, N}$ & $W_{L}^{-\,\mu}$ & / &W8 & $\left(T_{4}^{i}\right)_{2}$ & $W_{R}^{+\,\mu}$ & $W_{R}^{-\, N}$ & /\tabularnewline
\hline 
W9 & $\left(T_{4}^{i}\right)_{2}$ & $\left(\xi_{1}^{j}\right)_{21}$ & $W_{R}^{-\, N}$ & $W_{R}^{+\, N^{\prime}}$ &  &  &  &  & \tabularnewline
\hline 
\end{tabular}\end{center}

\caption{Possible field configuration inside the loop of the non-abelian diagram
topologies W1-W8\label{tab:Possible-field-configurationNA}. A capital roman index  on a gauge field indicates that 
both the vector and the scalar component are valid options, a small Greek index shows that only the vector 
components may propagate.}
\end{table}

\begin{figure}
\begin{minipage}{0.45\textwidth}
\begin{center}
 \includegraphics[width=0.95\textwidth]{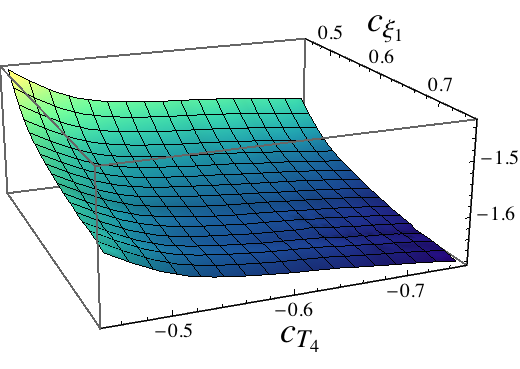}
\end{center}

\end{minipage}
\begin{minipage}{0.45\textwidth}
\includegraphics[width=0.95\textwidth]{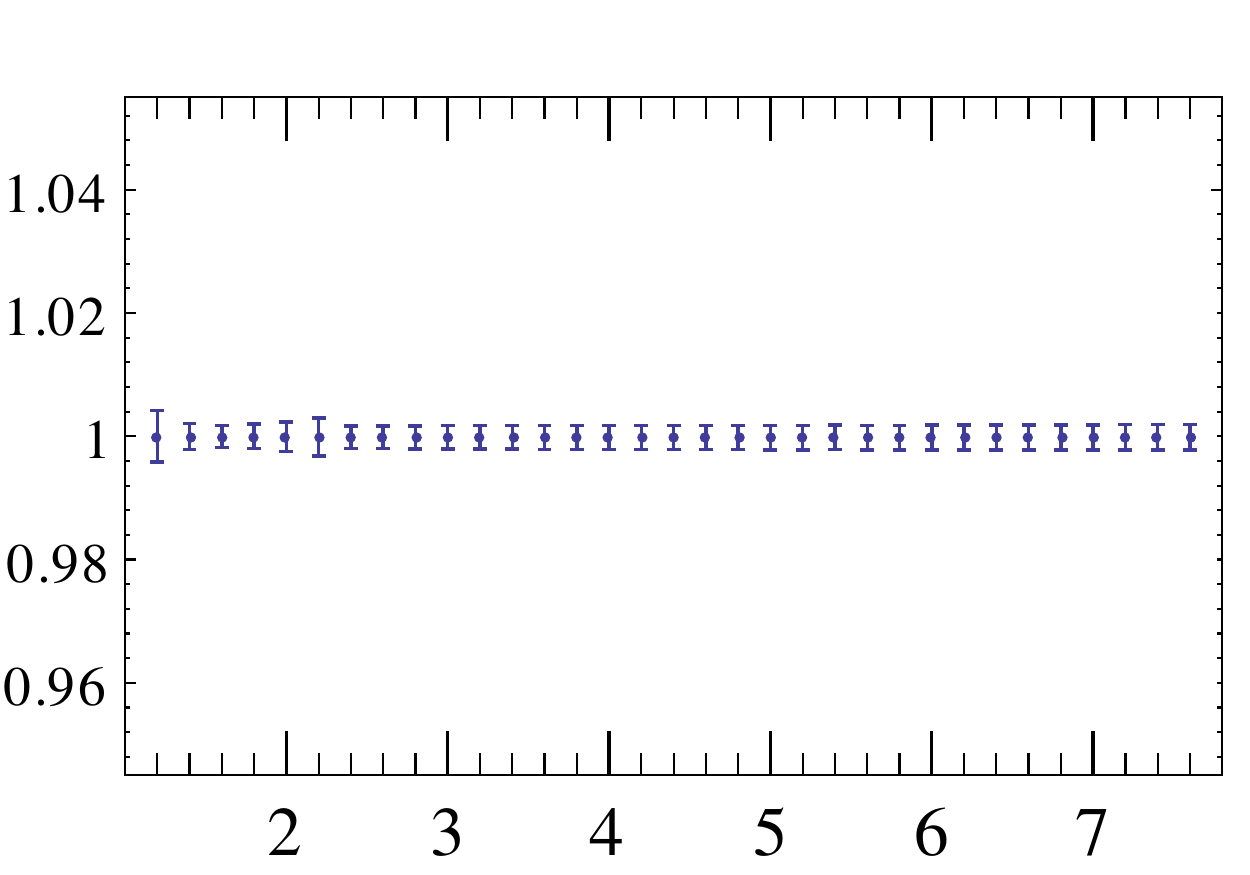}
\end{minipage}
\caption{\label{fig:MassGaugeDep}{\it left panel:} $10^8\cdot \mathcal{A}_{ij}$ as a function of the 5D mass parameters 
                                          $c_{{\xi_1}_i}$ and $c_{{T_4}_j}$ for  $T=1000\,\rm GeV$, 
                                          $k=2.44\cdot 10^{18}\rm GeV$.
                         {\it right panel:}   Residual dependence  of  $a_{ij}$ on the gauge parameter $\xi$ normalised on the value of $a_{ij}$ for $\xi=2$
                           (the error bars indicate the numerical uncertainties as estimated by our integration routines {\bf [J]}).            
                                          }
\end{figure}

Obviously, we cannot give the Wilson coefficient of the dipole operators in a closed analytical expression.
However, it is possible to give a graphical representation 
of its dependence on the 5D parameters. To this end, we rewrite $a^{\rm g}_{ij}$, the gauge contribution to 
$a_{ij}$, in the following way
\begin{align}
\label{eq:WilsonCoeffReduced}
a^{\rm g}_{ij}= \lambda^{d}_{ij} \frac{T^3}{k^3}f_{L_i}^{(0)}(1/T) g_{E_j}^{(0)}(1/T) \mathcal{A}_{ij}\;,
\end{align}
where no summation over repeated indices is performed.
The prefactors in \eqref{eq:WilsonCoeffReduced} have been chosen such that
$\mathcal{A}_{ij}$ is given by $a^{\rm g}_{ij}$ with each matrix element rescaled 
by the corresponding entry in the lepton mass matrix. 
 $\mathcal{A}_{ij}$ is a function of only $T$, the 5D mass parameters
$c_{{T_4}_i}$ and $c_{{\xi_1}_j}$ and the Planck scale $k$. In particular, we can interpret
$\mathcal{A}_{ij}$ as a measure for the misalignment of the Wilson coefficient $a^{\rm g}_{ij}$ relative to the mass matrix {\it before}
rotation into the mass eigenbasis. If, for fixed $T$ and $k$, $\mathcal{A}_{ij}$ is a constant in mass parameter space
there is no misalignment to leading order in $v^2/T^2$ and the dipole operator is not lepton-flavour violating.
Conversely, a strong dependence on the c-parameters indicates sizable FCNCs after EWSB.
Figure \ref{fig:MassGaugeDep} (left panel)  shows our result for $\mathcal{A}_{ij}$.  
The dependence on the 5D bulk masses is mild ($\pm 10\%$ for the typical range of 
mass parameters). Hence, we can expect the anomalous magnetic moment to be basically independent of the 
5D Yukawa structure and the dipole operator to induce only small flavour-violating transitions.
It should be noted that the relative variation of $\mathcal{A}_{ij}$ is by roughly a factor of two 
larger than in the minimal model, while its magnitude changed by a factor of three.

To check our result and to obtain an independent estimate on our numerical precision
one can study the residual dependence on the 5D gauge parameter $\xi$. 
The right panel in figure \ref{fig:MassGaugeDep} shows $a^{\rm g}_{ij}$ as a function of $\xi$
normalised on the value for $\xi=2$ for the parameter set $(T,c_L,c_E)=(1\; \rm TeV, 0.5478,-0.5478)$.  
The relative variation of our results with the gauge
parameter is below 1 per mille. This is smaller than the typical error estimate ($\sim 2$\textperthousand) provided by the 
numerical integration routine. In Feynman gauge ($\xi=1$) the numerical uncertainties are even smaller (below $1$\textperthousand), 
as the integrand takes a particularly simple form.  

\subsection{Higgs Contributions}
\label{sec:higgs}
The previously discussed gauge contributions are, at least to leading order in the couplings, 
not sensitive as to how the Higgs localisation is regularised \cite{Beneke:2012ie}. This is not the case
for contributions that are not linear in the Yukawa couplings. In the following, we use the simple regularisation
\begin{align}
\label{eq:HiggsRegulator}
\delta(z-1/T)=\lim_{\delta\to 0} \frac{T}{\delta} \Theta(z-\frac{1-\delta}{T}).
\end{align} 
To fix our model we also need to specify in which order the regulator $\delta$ 
and the regulator for the loop integrals, e.g.,~the dimensional regulator $\epsilon$
or a cut-off are taken to their physical values. It should be noted that the Higgs  
profile \eqref{eq:HiggsRegulator} introduces the new scale $T/\delta$ into 
the theory. All diagrams that are sensitive to the precise way in which the Higgs is 
localised have to be analysed with this new momentum region in mind, see \cite{Beneke:2012ie, Beneke:2014sta}
for detailed examples.

\begin{figure}
\begin{minipage}{0.03\textwidth}
{\footnotesize \mbox{ }\hspace{0.25cm} H1}\\
\mbox{ }
\end{minipage}
\begin{minipage}{0.205\textwidth}
\includegraphics[width=1.0\textwidth]{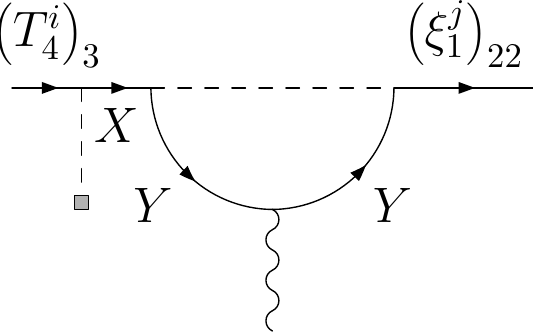}
\end{minipage}
\begin{minipage}{0.03\textwidth}
{\footnotesize \mbox{ }\hspace{0.25cm} H2}\\
\mbox{ }
\end{minipage}
\begin{minipage}{0.205\textwidth}
\includegraphics[width=1.0\textwidth]{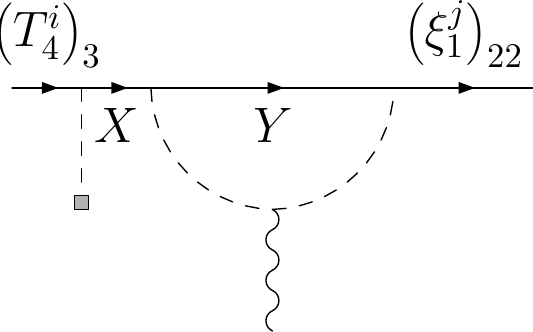}
\end{minipage}
\begin{minipage}{0.03\textwidth}
{\footnotesize \hspace{0.25cm} H3}\\
\mbox{ }
\end{minipage}
\begin{minipage}{0.205\textwidth}
\includegraphics[width=1.0\textwidth]{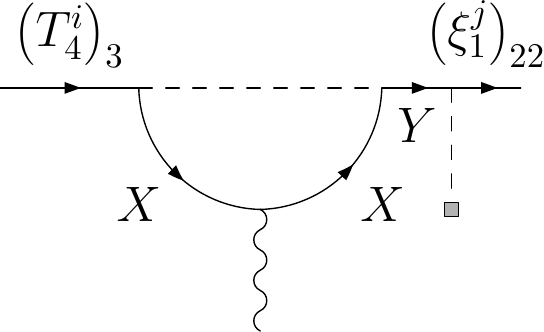}
\end{minipage}
\begin{minipage}{0.03\textwidth}
{\footnotesize \hspace{0.25cm} H4}\\
\mbox{ }
\end{minipage}
\begin{minipage}{0.205\textwidth}
\includegraphics[width=1.0\textwidth]{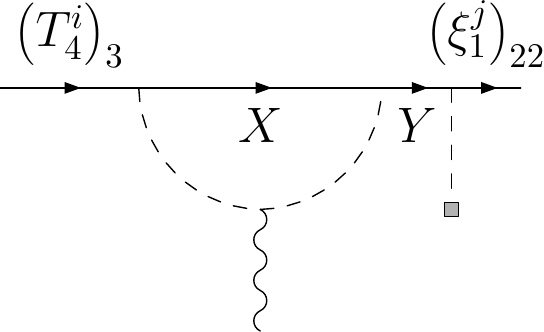}
\end{minipage}
\caption{\label{fig_collectedHiggstopologies} All topologies with internal Higgs bosons that contribute to the matching of
the dipole operator Wilson coefficient at one loop. $X$ and $Y$ label the species of the internal propagators.
The only assignments that do not cause the diagrams to trivially
vanish are: $X=\xi_1$ for all diagrams and $Y\in\{T_3,T_4\}$ for H1, H2 and H4 as well as $Y\in\{\xi_2,T_3,T_4\}$ for H3.}
\end{figure}
The diagrams contributing to the dipole operators are shown in figure \ref{fig_collectedHiggstopologies}.
Their short-distance part can be evaluated analytically for the choice \eqref{eq:HiggsRegulator}. 
We use dimensional regularisation for the $(4-2\epsilon)$ dimensional momentum integrals; 
the integrals over the fifth dimension are then trivial. 

If we choose to keep the $\delta$ regulator finite until all other regulators
have been removed from the theory, we obtain
\begin{align}\label{eq:Higgsresults}
a^H_{ij}=&\phantom{+} \frac{T^{11}}{2 k^{12}}  \frac{Q_\mu e}{192\pi^2} \left[
  f^{(0)}_{L_i}(1/T) Y_{ik}^{d} g^{(0)}_{E_k}(1/T)^2 Y^{\dagger,d}_{kh}F(c_{{\xi_1}_h})  Y_{hj}^{d} g^{(0)}_{E_j}(1/T) \right. \nn \\
& \left.    \qquad       \qquad    \qquad +2   f^{(0)}_{L_i}(1/T)Y^u_{ik}   F(c_{{\xi_2}_k})   
 Y^{u,\dagger}_{kh} f^{(0)}_{L_h}(1/T)^2    Y_{hj}^{d} g^{(0)}_{E_j}(1/T) \right]
 \nn \\
 & +  \frac{T^3}{k^4} \frac{Q_\mu e}{192 \pi^2}    \left[  f^{(0)}_{L_i}(1/T) [Y^dY^{\dagger,d} Y^d]_{ij} g^{(0)}_{E_j}(1/T)        
      -  f^{(0)}_{L_i}(1/T) [Y^uY^{u,\dagger} Y^d]_{ij} g^{(0)}_{E_j}(1/T)   \right]     
\end{align} 
with 
\begin{align}
F(c)=  
  \frac{k^4 ((1+2c) + (3-2c)\epsilon^{2-4c} -\epsilon^{1-2c}(3+4c-4c^2  
            )   )  }
 {(4c^2-4c -3) T^5 (1-\epsilon^{1-2c})^2 }\;
 \end{align}
for the Higgs contributions to the dipole operator.
$Y^{d/u}=\lambda^{d/u}_{(5D)} k$ are dimensionless 5D Yukawa matrices  
and we dropped terms that are suppressed by powers of the small ratio $\epsilon$. 
Note the setting the Yukawa $Y^u$ that 
arises via the $\bar\xi_1 \Phi \xi_2$ term in the Lagrangian to zero reproduces the 
result from the minimal model. This is due to an accidental cancellation of the various
new diagrams. 

If we choose to send the width of the Higgs to zero {\it before}
the regulator of the 4D loop integral, the last line in \eqref{eq:Higgsresults}  
 vanishes identically, see also \cite{Beneke:2012ie}. In this situation the 
dipole operator induced by Higgs exchanges is typically much smaller than in the case
with a reversed order of the limits as the third line in \eqref{eq:Higgsresults} 
is numerically dominant in almost all points of the parameter space.

As in case of the gauge contributions, the result for the  dipole coefficient is scheme dependent.
It depends on the treatment of $\gamma_5$ in $4-2 \varepsilon$  dimensions; 
\eqref{eq:Higgsresults} corresponds to naive dimensional regularisation (NDR). 
The scheme dependence is only cancelled when including the 
contributions of the operators $  \Phi^\dagger \Phi\; \bar L_i \Phi E_j $ and 
$\Phi^\dagger i\overleftrightarrow{D}_\mu \Phi  \;\bar E_i \gamma^\mu E_j $ 
whose contributions vanish to our order in the expansion in NDR, but are
finite in other schemes.

\section{Numerical Result}

With the Wilson coefficients at hand, we can now discuss the modification 
of the anomalous magnetic moment in the custodially protected RS model.
The main input parameters are the 5D Yukawa matrices and the 5D masses.
As the theory has to reproduce the SM lepton sector in the low energy limit
we are provided with the additional constraints via lepton masses and mixing.
For simplicity, we only require that the charged lepton masses are reproduced, 
all (Dirac) neutrino masses are below $0.1 \;\rm eV$ and that their mass splitting 
does not violate the bounds from neutrino oscillation; we do not require that the 
PMNS matrix is reproduced. It should be noted that 
the dependence on $c_{\xi_2}^i$ (which only enters in terms with at least three 
Yukawa factors) is quite small. 
\begin{figure}
\begin{minipage}{0.47\textwidth}
\includegraphics[width=0.95\textwidth]{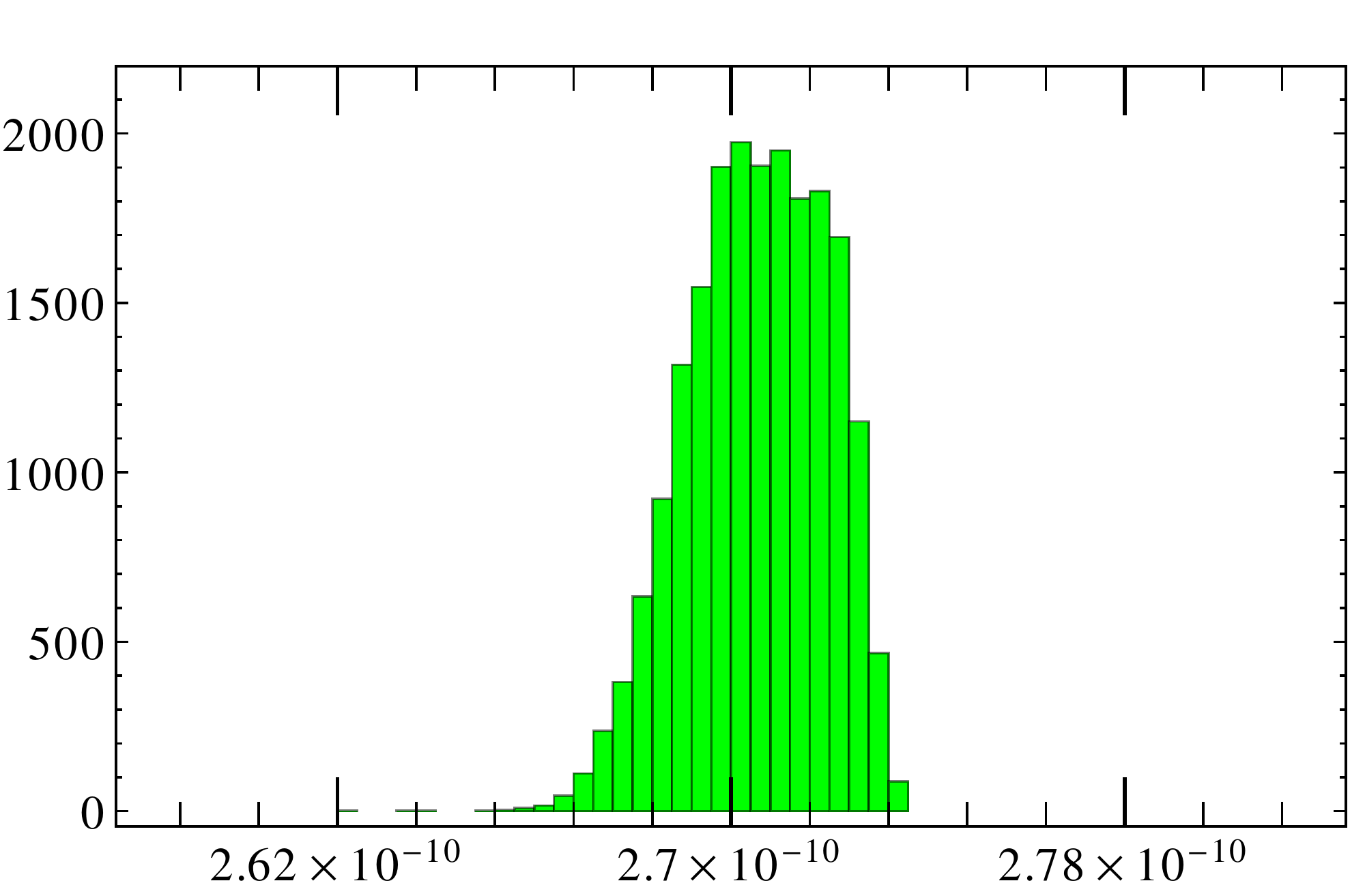}
\end{minipage}
\begin{minipage}{0.47\textwidth}
\includegraphics[width=0.95\textwidth]{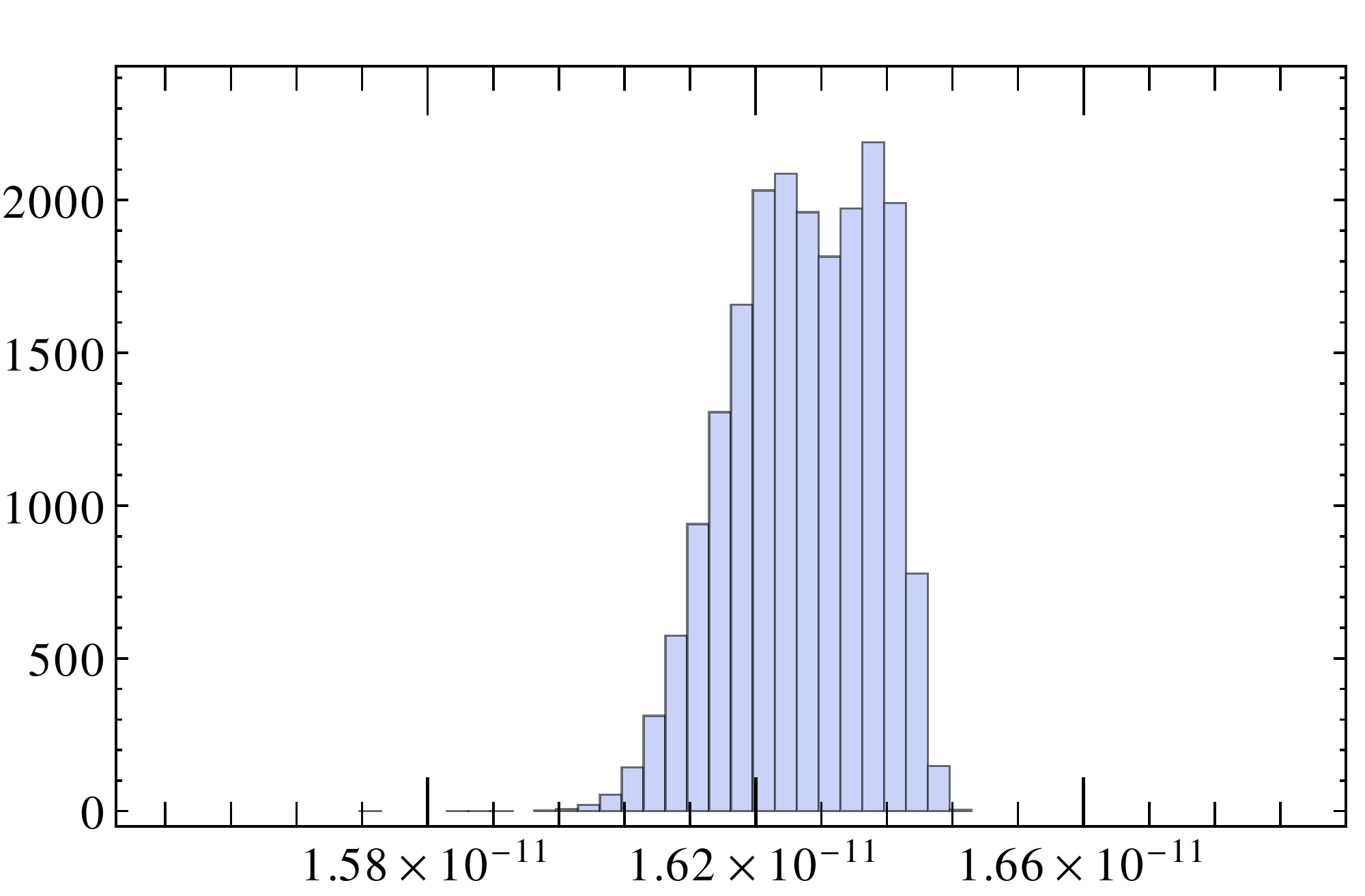}
\end{minipage}
\caption{\label{fig:plotsforGm2}{\it left panel:} Histogram of the contribution of the gauge diagrams $\Delta a^{\rm g}$
to the anomalous magnetic moment $a_\mu$ for fixed $T=1\;\rm TeV$. {\it right panel:} Same for $T=4\;\rm TeV$. }
\end{figure}
We randomly generate parameter sets that pass all constraints
and determine $a_\mu$ for each set. Following the general spirit of the RS 
model our Yukawas are anarchic, i.e.~each matrix element is very roughly of modulus one 
and has a random phase. In practice, we allow for values from $1/10$ to $10$ for the absolute value
and use a flat distribution for its logarithm.

The gauge contributions are expected to be virtually independent of our parameter choice,
as $\mathcal{A}_{ij}$, cf.~\eqref{eq:WilsonCoeffReduced}, is approximately mass-parameter independent.
The left panel of figure \ref{fig:plotsforGm2} shows the result for $\Delta a^{\rm g}_\mu$ for 
a fixed value of $T=1\;\rm TeV$. The result for $T=4\;\rm TeV$ is shown in the right panel. The histograms are generated from each
$10^5$ parameter sets.
The distribution is centred around $2.72 \cdot 10^{-10}$ for $T=1\;\rm TeV$ while 
$T=4\;\rm TeV$ lead to a central value of $1.63 \cdot 10^{-11}$. This is in line with the 
typical scaling $\Delta a_\mu^{\rm gauge}\propto \frac{1}{T^2}\times\ln k/T $.
The model-independent gauge contribution
to $g_\mu-2$ in the custodially protected RS model can thus be reliably estimated via 
\begin{align}\label{eq:TotalGaugeContribution}
 \Delta a_{\mu}^{\rm g}\approx 2.72 \times 10^{-10} \left(\frac{1\;\rm TeV}{T} \right)^2
\end{align}
for any phenomenologically relevant value of $T$.

Comparing to the result in the minimal model \cite{Beneke:2012ie}, 
$\Delta a_\mu^{\text{min}}\approx 0.88\cdot 10^{-10}(1 \;\rm TeV)^2/T^2 $
we see the minimal model gives a correction to the anomalous magnetic moment 
that is roughly a factor of $3$ smaller, while the T dependence is, as expected, the same.
Despite the significant enhancement compared to the minimal model,
 more realistic choices of $T > 2000 \;\rm GeV$ (which corresponds to KK masses larger than $4.7\; \rm TeV$)
 only gives an enhancement to $a_\mu$ of at best  
 $6.8 \cdot 10^{-11}$. The difference between the current experimental value and the SM prediction 
 for the anomalous magnetic moment of the muon is given by \cite{Beringer:1900zz}
\begin{align}\label{eq:Gminus2Tension}
a_\mu^{\rm exp}-a^{\rm SM}_\mu=287(63)(49)\times 10^{-11}
\end{align} 
where theory and experimental uncertainties are given separately. Thus, the 
gauge contribution $\Delta a_\mu^{\rm g}$ to $a_\mu$ alone is too small to be noticed in experiments.

The effect of the modified W/Z coupling $\Delta a^{ZW}_{\mu}$ is not included in the above numbers.
For mass parameters $|c_{L/E}|>0.55$ it is given by
\begin{align}
 \Delta a^{ZW}_{\mu}\approx -0.46\cdot 10^{-11} \left(\frac{1\;{\rm TeV}}{T}\right)^{2}
\end{align}
and is, for general 5D masses of the order of $\text{few}\times 10^{-12} \frac{1\;{\rm TeV}^2}{T^2}$
in both the minimal and custodially protected model. This is negligible for the custodially protected and
a $\sim 5\%$ correction in the minimal model.

The Higgs contributions are strongly dependent on the model parameters, especially the Yukawa
matrices. So general statements as in the case of the gauge contribution are not feasible.
However, it is worthwhile to study the effect of the Higgs exchange in several illustrative 
scenarios. As only one of the two Higgs localisations discussed in section \ref{sec:higgs}
gives rise to sizable corrections to the dipole operators the following discussion will focus 
on this case only.
 
\begin{figure}
\begin{minipage}{0.47\textwidth}
   \includegraphics[width=1\textwidth]{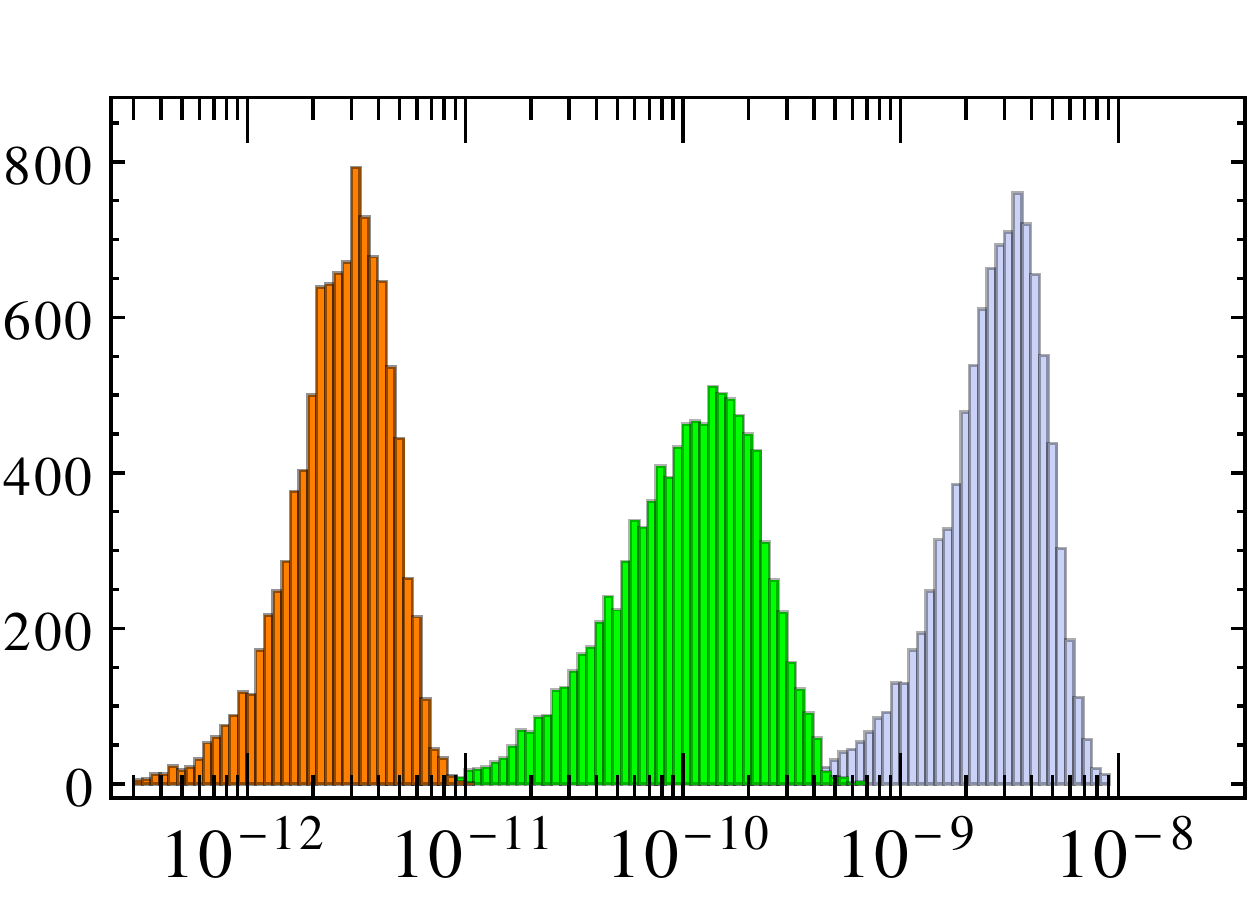}
\end{minipage}
\begin{minipage}{0.47\textwidth}
   \includegraphics[width=1\textwidth]{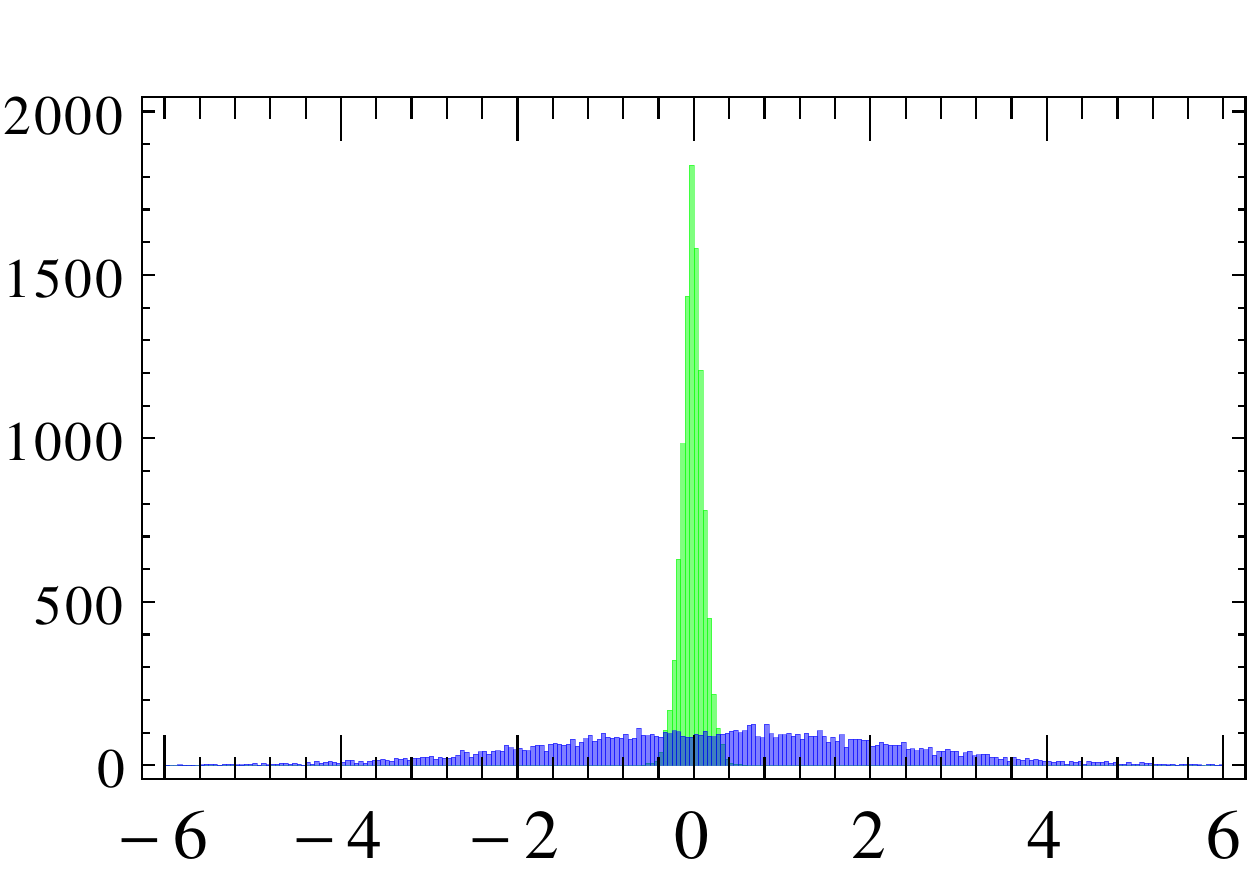}
\end{minipage}
\caption{\label{fig:HiggsMINplotsforGm2} {\it left panel:} 
Higgs contribution to $g_\mu-2$ for different  
average Yukawa sizes in the minimal model (see text for details).  The red (light grey) histogram corresponds to $|Y|\in(1/10,1/3)$, the 
green(grey) one  to $|Y|\in(1/3,3)$ and blue (dark grey) to $|Y|\in(3,10)$.
 We use $T=1\;{\rm TeV}$ everywhere. {\it right panel:} 
As left panel but for the custodially protected model and without the histogram  for small
Yukawa matrices. The x-axis uses a linear scale in units of $10^{-9}$.  }
\end{figure}

Let us first go back to the minimal RS model which was already discussed in \cite{Beneke:2012ie}.
In this case we only need to consider the first term in each 
square bracket in \eqref{eq:Higgsresults}. Obviously, the contribution to $g_\mu-2$ 
will increase with the magnitude of the Yukawa matrix (in the minimal case 
there is only one lepton Yukawa). To quantify this statement we 
study the shift of $(g-2)_\mu$ due to the dipole Wilson coefficient $a^H$ 
for three hypothetical cases: the Yukawa entries are each 
in the range $(1/10,1/3)$, $(1/3,3)$ or $(3,10)$, $T$ is fixed to $1\;\rm TeV$ and we generate
$10^4$ random data sets for each scenario. 
Figure \ref{fig:HiggsMINplotsforGm2} (left panel) shows the result
for the different Yukawa ranges using a logarithmic scale for the abscissa.
One can see that the central values of the histograms 
scale with the square of the corresponding average Yukawa size. This was to be expected from 
\eqref{eq:Higgsresults} as the product of zero-mode profiles compensates for one
Yukawa factor provided the Yukawa matrices themselves do not carry a strong hierarchy. 
As each of the  distributions is spread out over more than an order of magnitude it 
is not possible to make quantitative statements without a detailed knowledge of the 
Yukawa matrices. 
We also find that $a^H$ favours a positive contribution 
to $(g-2)_\mu$ if one constrains the Yukawas as described.
Here the logarithmic scale on the x--axis is slightly misleading:
it illustrates the scaling with the Yukawa size but misses a short tail 
in the negative region. Nonetheless, the contributions are predominantly 
positive. This is interesting as the Higgs contribution is then aligned 
with the gauge contribution: both reduce the
difference between theory and experiment \eqref{eq:Gminus2Tension}.
We can use the current limits on $g_\mu-2$ to give a rough bound on the ratio 
$\frac{\langle YY^\dagger \rangle}{T^2}$. The bound is, in a sense, maximally weak,
as the preference for a positive sign forces us to consider $\Delta a_\mu^{RS}<6\cdot 10^{-9}$
as an upper bound. Thus the constraining power of $g_\mu-2$ for the lepton Yukawa
sector is weaker then Higgs production \cite{Malm:2013jia}
is for the quark sector even though both are sensitive to the same
ratio $\frac{\langle YY^\dagger \rangle}{T^2}$.

Only average Yukawa entries of at least $3$ would allow for a correction that is sizable enough to 
remove the current tension. However, such large values would, assuming Yukawa anarchy,
also effect other observables. We also find that the general  T-dependence is 
in agreement with the expected $1/T^2$ behaviour from power-counting. 

Next, we turn to the custodially protected model. We now need to include the term with the 
novel Yukawa $Y^u$ in \eqref{eq:Higgsresults}. If both Yukawas were the same, the different 
contributions would cancel in the dominant term and the total contribution would be small
compared to $\Delta a^{\rm g}_\mu$. However, there is generally no argument that can be used 
to fix the relative size of the Yukawa matrices. In the following we assume that the 
two matrices have a common entry size. The right panel of figure 
\ref{fig:HiggsMINplotsforGm2} shows the Higgs contribution
to the anomalous moment. We only show the two cases of large and intermediate Yukawa 
entries; on a linear scale the histogram for small Yukawas entries is too narrowly 
centred around zero to be visible.

As in the minimal scenario we find potentially very large corrections
to $g_\mu-2$ for $T=1 \; \rm TeV$. The effect for different choices of $T$
can again be obtain by making use of the overall $1/T^2$ scaling. 
The preference for a positive contribution that is present in 
the minimal model is not observed. 
However, we have to be careful if we want to make statements
about possible size of the Higgs contributions.   
It is well-known that the dipole operator enters not only 
in the anomalous magnetic moment but also various other 
observables in lepton flavour physics; notably the
it gives a potentially large contribution to $\mu \to e \gamma$.
Note, that the gauge contribution to $g_\mu-2$ is basically independent of 
the Yukawa structure and the contributions to off-diagonal 
flavour-changing transitions are suppressed. Hence, 
experimental bounds on FCNCs will be less important there.

A complete analysis of lepton flavour observables in the RS model will be 
presented elsewhere. In the following we 
just illustrate their potential impact by considering the dipole operator 
contribution to $\mu\to e\gamma$ alone. Since $\mu \to e \gamma$ can be mediated by the coefficients
$\alpha_{12}$ and $\alpha_{21}$ in the same way as $(g-2)_\mu$ is mediated by $\alpha_{22}$, 
it is straightforward to estimate the branching fraction.
A naive estimate that assumes that the dipole operators are essentially structureless in
flavour space was already given in \cite{Beneke:2012ie,Beneke:2014sta}. Under this assumptions the 
bounds on FCNCs all but eliminate the Higgs contribution to $g_\mu-2$.
However, especially in the custodial model where we observe two competing contributions
due to the presence of an additional Yukawa matrix, this picture may be too simple. 
In the following we will  use the upper bound of $5.7\times10^{−13} $ for the $\mu \to e \gamma$
branching ratio \cite{Adam:2013mnn}. We again generate random 
parameter points and require that the dipole operator contribution 
to $\mu \to e \gamma$ alone does not violate this bound.
 
 \begin{figure}[t]\begin{center}
\begin{minipage}{0.57\textwidth}
\includegraphics[width=1\textwidth]{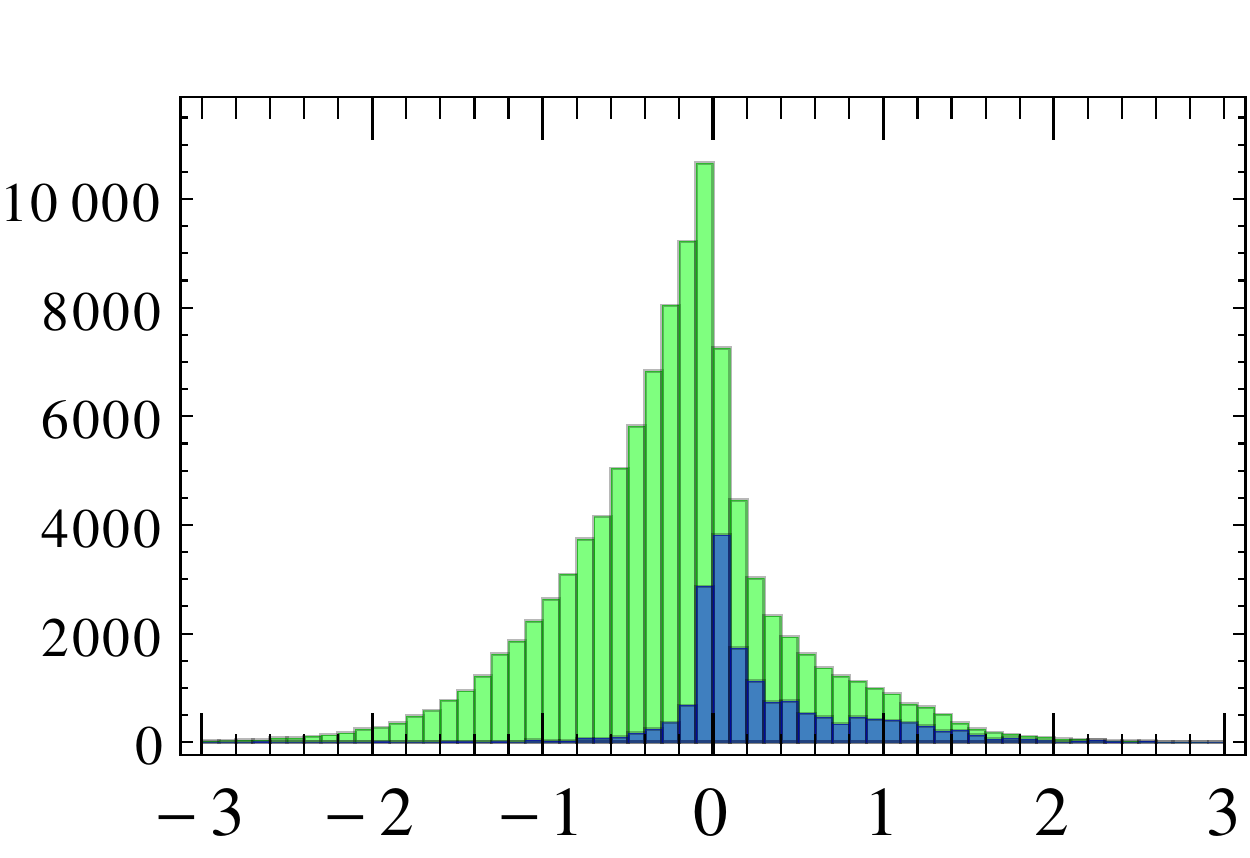}
\end{minipage}\end{center}
\caption{\label{fig:HiggsMEGplotsforGm2}
Histogram of the Higgs contribution 
to $g_\mu-2$ in units of $10^{-10}$  for $T=4\;\rm TeV$. 
Without (green/light grey) and with (blue/dark grey) taking the $\mu \to e \gamma$
branching ratio into consideration. 
The height of the blue histogram  has been increased by a factor of ten. 
}
\end{figure}
In figure \ref{fig:HiggsMEGplotsforGm2}
we show effect of $a^H$ on $g_\mu-2$ with (in blue) 
and without (in green) the  $\mu \to e \gamma$ bound for $T=4\;\rm TeV$. 
The Yukawa matrix elements where allowed to take values in the interval $(1/10,10)$. 
The green (light grey) histogram consists of $10^5$ parameter sets.
while only $\approx 2000$ sets survive the MEG bound.
After taking the bound into account, we find a 
preference for positive contributions $a^H_\mu$ as was already the case in 
the minimal model.
The asymmetry arises because the terms with a $Y^u$ factor in \eqref{eq:Higgsresults} tend to 
generate larger off-diagonal elements in $a_{ij}^H$ than the terms with only one type of Yukawa 
matrix. Hence, the second term in the third line of \eqref{eq:Higgsresults} is affected 
more strongly than the predominantly positive first term.   
For $T=1\;\rm TeV$ the $\mu \to e \gamma$ bound has an even more pronounced effect:
if one e.g.~restricts the Yukawa entries to the interval $(3,10)$ no parameter sets
that pass the MEG bound were found in a random sample of $10^5$ sets and only
one in $4000$ parameter sets with Yukawas in the interval $(0.1,10)$ passed the constraint for $T=1\;\rm TeV$.
We have to stress, that the random scans through parameter space can 
only give indications to which extent the contributions to $g_\mu-2$
are constrained. For a more definite statements a comprehensive 
analysis of lepton (flavour) observables in the RS model would be 
necessary.

\section{Conclusions}
We presented the computation of the leading effects of the 
custodially protected Randall-Sundrum model on the anomalous
magnetic moment of the muon. To this end we extended the techniques
developed in \cite{Beneke:2012ie} for the case of the minimal RS model. We find that the 
contribution to $a_\mu$ mediated by an exchange of a virtual gauge boson
is given by
\begin{align}
 \Delta a_\mu =  2.72 \cdot 10^{-10}\left(\frac{1 \;\rm TeV}{T}\right)^2.
\end{align}
This number is essentially independent of the  model parameters in the 
lepton sector and by more that a factor of three larger than
the corresponding result in the minimal RS scenario. Despite the 
enhancement the effect is still insufficient to reconcile theory and 
experiment.

The effect of diagrams with a virtual Higgs boson can be determined 
analytically, but requires a precise specification of the
localisation of the Higgs near the IR brane.  The result \eqref{eq:Higgsresults}
is valid for the so-called narrow bulk Higgs (in the language of \cite{Malm:2013jia}).
Since the Higgs contribution exhibits by definition a strong dependence on 
the model parameters in the Yukawa sector, one can use e.g.~$\mu \to e \gamma$
to limit the magnitude of the dipole operators, however, 
such bound can always be circumvented by imposing 
a specific flavour structure already in the 5D Lagrangian. 
Without bounds from lepton flavour physics
the Higgs contribution can reach values as large as
\begin{align}
  |\Delta a^H_\mu| \lessapprox  {\text{few}} \times 10^{-9}  \times \left(\frac{1 \;\rm TeV}{T}\right)^2 
\end{align}
if the entries of the dimensionless 5D Yukawa matrices are allowed to have magnitudes 
of up to $10$. In the minimal model the contribution is predominantly positive, in the custodially protected 
model the sign is undetermined.  While the inclusion of the current bound on $\mu \to e \gamma$
seems introduce a preference for a positive shift in $g_\mu-2$, this has 
to be studied in more detail and  with a larger set of potentially 
sensitive lepton flavour observables to make a definite statement. 
 
\paragraph{Acknowledgements:} We thank M.~Beneke for many 
  helpful discussions. The work of P.M.~ is supported in part by the Gottfried Wilhelm Leibniz programme
 of the Deutsche Forschungsgemeinschaft (DFG).
  The work J.R.~is supported by STFC UK.
  The Feynman diagrams were drawn with the help of
  Axodraw \cite{Vermaseren:1994je} and JaxoDraw \cite{Binosi:2003yf}.

\end{document}